\pdfoutput=1

\documentclass[aip,jcp,reprint,groupedaddress,footinbib,citeautoscript]{revtex4-1}
\usepackage{graphicx} %
\usepackage{amsmath}
\usepackage{mathtools} %
\usepackage{bm}
\usepackage{amssymb} %

\usepackage{hyperref}
\hypersetup{colorlinks=true,linkcolor=blue,citecolor=blue,urlcolor=blue}

\usepackage{braket} %

\newcommand{\eu}{\mathrm{e}^}
\newcommand{\iu}{\ensuremath{\mathrm{i}}}
\newcommand{\rmd}{\mathrm{d}}
\newcommand{\thalf}{{\ensuremath{\tfrac{1}{2}}}} %
\newcommand{\half}{{\ensuremath{\frac{1}{2}}}}

\newcommand{\op}[1]{\ensuremath{\hat{#1}}}

\renewcommand{\mathbf}[1]{\bm{#1}}

\DeclareMathOperator{\Tr}{Tr}

\newcommand{\T}{{\mathstrut\top}}

\newcommand{\pder}[3][]{\frac{\partial^{#1}{#2}}{\partial{#3}^{#1}}}

\newcommand{\eqn}[1]{Eq.\,(\ref{#1})}

\newcommand{\fig}[1]{Fig.\,\ref{fig:#1}}
\newcommand{\Fig}[1]{Figure \ref{fig:#1}}
\newcommand{\tref}[1]{Table \ref{tab:#1}}

\newcommand{\Ref}[1]{Ref.~\onlinecite{#1}}

\graphicspath{{fig/}}

\begin{document}

\title{Effects of tunnelling and asymmetry for system-bath models of electron transfer}

\author{Johann Mattiat}
\author{Jeremy O. Richardson}
\email{jeremy.richardson@phys.chem.ethz.ch}
\affiliation{Laboratory of Physical Chemistry, ETH Zurich, 8093 Zurich, Switzerland}

\date{\today}

\begin{abstract}
\noindent
We apply the newly derived nonadiabatic golden-rule instanton theory
to asymmetric models describing electron-transfer in solution.
The models go beyond the usual spin-boson description
and have anharmonic free-energy surfaces
with different values for the reactant and product reorganization energies.
The instanton method
gives an excellent description of the behaviour of the rate constant with respect to asymmetry for the whole range studied.
We derive a general formula for an asymmetric version of Marcus theory
based on the classical limit of the instanton
and find that this gives significant corrections to the
standard Marcus theory.
A scheme is given to compute this rate based only on equilibrium simulations.
We also compare the rate constants obtained by the instanton method with its classical limit to study the effect of tunnelling and other quantum nuclear effects.
These quantum effects can increase the rate constant by orders of magnitude.
\end{abstract}

\maketitle

\section{Introduction}

Chemical reactions involving electron transfer (ET) occur in many different environments, from redox reactions to photosynthesis and the harvesting of light in solar cells \cite{Marcus1993review}.
In the simplest ET reactions,
the charge is transferred from a donor or acceptor,
which can be as small as solvated ions \cite{Kuharski1988Fe3e}, or as large as protein complexes \cite{Blumberger2008ET}.
Thus there are at least two important electronic states involved in the reaction
and typically the Born-Oppenheimer approximation breaks down,
making it necessary to consider nonadiabatic dynamics to describe and predict the rate of the process.

In many cases, the rate can be considered to be in the golden-rule limit \cite{ChandlerET},
that is where the coupling, $\Delta$, between the electronic states is assumed to be weak
and the ET itself is the bottleneck to the reaction.
Fermi's golden-rule thus provides an acceptable formula for the exact rate constant of the process. \cite{Zwanzig}
It is obtained from perturbation theory with a lowest-order expansion in the coupling giving a rate proportional to $\Delta^2$.

Nonetheless as the eigenstates of the full Hamiltonian are generally not known,
approximations to Fermi's golden-rule formula have to be made.
Several theories have been proposed to tackle this kind of problem,
most famously by Marcus \cite{Marcus1956ET,Marcus1960ET,Marcus1964review,Marcus1985review,Marcus1993review}.
Its simple form and wide range of applications
make Marcus theory a standard approach to treat ET\@.
The theory is derived by applying a classical transition-state theory approximation to Fermi's golden rule
and making a number of assumptions about the shapes of the free-energy curves involved.

The standard assumption is that all nuclear degrees of freedom obey Gaussian statistics,
leading to parabolic free-energy curves along the reaction coordinate.
This describes the orientation of the nuclear coordinates,
otherwise known as the environment,
and is defined as the instantaneous vertical energy gap
between the two electronic states involved.
The ET rate can then be expressed in terms of the reorganization energy $\lambda$.
This is defined as the free energy that is required to change the reaction coordinate %
from its value which minimizes the reactant free energy to its value which minimizes the product free energy
without changing electronic state.
By construction, the reactant and product free-energy curves have the same curvature and reorganization energies even for a biased system. \cite{Tachiya1989ET}
The standard Marcus theory rate constant is given by \cite{Marcus1985review}
\begin{equation} \label{Marcus}
	k_\text{MT} = \frac{\Delta^2}{\hbar} \sqrt\frac{\pi\beta}{\lambda} \, \eu{- \beta \frac{(\lambda - \epsilon)^2}{4 \lambda}},
\end{equation}
where $\beta=(k_\mathrm{B}T)^{-1}$ and $\epsilon$ is the difference between the minimum free energy of the reactant and the minimum free energy of the product.
One of the main achievements of Marcus theory was the prediction of the behaviour of the rate
in the inverted regime, where $\epsilon>\lambda$, which was later confirmed by experiment \cite{Miller1984inverted}.

Much of the early work \cite{Marcus1956ET} was devoted to obtaining a formulation of the reorganization energy in terms of the dielectric continuum to describe the solvent.
However, in modern theoretical chemistry, atomistic molecular dynamics simulations
allow us to probe the microscopic quantities directly
and different techniques are required for computing the rate.
A number of early studies \cite{Hwang1987ET,King1990ET,Kuharski1988Fe3e,ChandlerET} show how the free-energy curves can be computed from statistical mechanics
and in many cases, the curves were found to be approximately parabolic in agreement with Marcus' assumptions.

There are however occasions where the parabolic assumption breaks down.
For asymmetric reactions, there is no reason why the environment around the reactant and product should behave in the same way
which leads to different reorganization energies for the reactant and product states.
Computer simulations have found extreme cases where the reactant and product reorganization energies differ by up to a factor of 2
both in ET between solvated ions \cite{Blumberger2006Ag} and between sites in proteins \cite{Krapf2012Marcus}.
This implies that the free-energy curves cannot be harmonic over all the reaction coordinate, although they may still be approximately harmonic around their equilibrium positions \cite{Tachiya1989ET},
and therefore that the standard Marcus theory formula cannot be applied.
The best solution to avoid this problem is to compute the free-energy curves directly from simulation \cite{ChandlerET};
using the energy gap as the reaction coordinate, the transition-state is found when the gap is zero.
This gives the activation energy directly rather than indirectly approximating it from the reorganization energy as in \eqn{Marcus}.
Nonetheless the simplicity of the standard formulation in terms of reorganization energies is attractive and many recent computational studies rely only on simulations of these quantities. \cite{Krapf2010Marcus}
We therefore propose in this work an asymmetric generalization of Marcus theory to treat such cases more accurately.

Quantum nuclear effects are also ignored by the standard Marcus theory.
These effects allow for tunnelling of the nuclear coordinates and are expected to lead to a speed-up of the rate.
One approach for including quantum effects into ET processes is to map the system onto a harmonic spin-boson model \cite{Garg1985spinboson}
and solve the resulting equations either using semiclassical approximations or numerically exactly \cite{UlstrupBook,Siders1981quantum,*Siders1981inverted,Bader1990golden,Topaler1996nonadiabatic,Wang2006flux}.
This approach however cannot take account of anharmonicity as described above,
although
certain generalized spin-boson systems can still be studied within these approaches \cite{Tang1994asym} and anharmonic effects treated within a perturbative approach \cite{Tang1994anharmonic}.

A method that promises to overcome both of the limitations
of Marcus theory discussed above
is semiclassical instanton theory \cite{Miller1975semiclassical,Coleman1977ImF,*Callan1977ImF,Affleck1981ImF,Benderskii,Andersson2009Hmethane,RPinst,Althorpe2011ImF,AdiabaticGreens,faraday,Kaestner2014review,Ceotto2012instanton},
which was recently extended to describe electron transfer in the nonadiabatic limit by one of us \cite{GoldenGreens,GoldenRPI}.
This approach is applicable to multidimensional anharmonic potential-energy surfaces and takes both zero-point energy and nuclear tunnelling into account.
Only simple numerical algorithms including a saddle-point optimization are required to apply the method and it is therefore computationally inexpensive.
Additionally, the classical limit of this instanton theory,
which will be formulated below,
can be compared to the classical results obtained from standard Marcus theory.

Alternatively, a path-integral Monte Carlo method \cite{Wolynes1987nonadiabatic,Zheng1989ET,*Zheng1991ET} can be employed to give an approximation to the quantum rate constant.
This was used by Chandler and coworkers to study the ferrous-ferric electron transfer \cite{Bader1990golden} and found a speed-up of a factor of 60 compared to classical approaches.
It is less computationally efficient than the instanton approach as it is necessary to sample a large number of path-integral configurations to achieve numerical convergence,
although has the advantage of being easier to apply to atomistic liquid systems.
For the system-bath model studied in this work, it is easy to show that the results of this approach will be equivalent to those of instanton theory.
However, in general for anharmonic systems the classical limit of this path-integral approach is not so clearly linked with transition-state theory and casts doubt on its applicability in all regimes \cite{nonoscillatory,GoldenGreens,GoldenRPI}.

There are also other effects which are neglected here,
and in order to describe certain problems it may be required to derive further extensions of the standard Marcus theory.
In particular, it is possible to go beyond the golden-rule limit
and compute rates for systems with stronger electronic couplings.
\cite{Rips1995ET,Muehlbacher2003spinboson,CasadoPascual2003ET,Barzykin2002ET,Spencer2016Faraday}
Semiclassical approximations for this have also been developed
including an instanton approach related to ours
\cite{Cao1995nonadiabatic,*Cao1997nonadiabatic,*Schwieters1998diabatic}
and Zhu-Nakamura theory.
\cite{Zhu1994ZN,*Zhu1995ZN,Zhao2004nonadiabatic}

In this work both classical and semiclassical methods will be applied to an asymmetric system-bath model with anharmonic free-energy curves
in order to explore the behaviour of the rates of various approaches with respect to both anharmonicity and tunnelling.

\section{Theory}

The Hamiltonian describing an ET process can be represented in the diabatic representation as
\begin{align}
	\hat{H} &= \hat{H_0} \ket{0} \bra{0} + \hat{H_1} \ket{1} \bra{1} + \Delta (\ket{0} \bra{1} + \ket{1} \bra{0}),
	\label{Hamiltonian}
\end{align}
where $\ket{0}$ and $\ket{1}$  are the electronic states of the reactant and product
which are coupled by $\Delta$.

Here we will assume that the Condon approximation holds, such that $\Delta$ is a constant.
However, 
the instanton approach and its classical limit
can be easily extended to describe a coordinate-dependent coupling $\Delta(\op{\mathsf{x}})$.
Within the steepest-descent approximation, the value of coupling used in the equations should simply be that of the hopping point, \textit{i.e.}\ $\Delta\equiv\Delta(\mathsf{x}^\ddagger)$.

The Hamiltonians $\hat{H}_0$ and $\hat{H}_1$ describe the nuclear degrees of freedom of each electronic state and are of the form
\begin{equation}
	\op{H}_n = \sum\limits_{j=1}^f \frac{\hat{p}_j^2}{2m} + V_n(\hat{\mathsf{x}}), \qquad n = \{0,1\},
	\label{eq:sub_Hamiltonian}
\end{equation}
where $\mathsf{x}=(x_1,\dots,x_f)$ are the nuclear coordinates and
the functions $V_n(\mathsf{x})$ describe the reactant and product potential-energy surfaces (PES) on which the nuclei move. 
We use re-weighted coordinates such that each degree of freedom has the same mass, $m$. %
The rates do not depend on the choice of this parameter.

\subsection{Instanton theory}

The derivation of the semiclassical instanton approximation
to the thermal rate
in the weak-coupling, golden-rule limit
is performed in a step-by-step manner in \Ref{GoldenGreens}.
This follows a procedure almost identical to that used to obtain a rigorous rate theory
in the adiabatic limit.
\cite{AdiabaticGreens,faraday}
It is thus related to the standard instanton formulas applicable when the Born-Oppenheimer approximation is valid.
\cite{Miller1975semiclassical,Coleman1977ImF,*Callan1977ImF,Affleck1981ImF,Benderskii,Andersson2009Hmethane,RPinst,Althorpe2011ImF,Kaestner2014review,Ceotto2012instanton} 
Here we present an equivalent derivation
in a more direct manner.

The flux correlation formulation \cite{Miller1974QTST,Miller1983rate} gives the exact rate constant, $k$, in the golden-rule limit as
\begin{align}
	k Z_0
	&= \frac{\Delta^2}{\hbar^2} \int_{-\infty}^\infty C^\tau(t) \, \rmd t,
	\label{exact}
\end{align}
where
\begin{align}
	C^\tau(t) &= \Tr \left[ \eu{\iu\op{H}_0(t+\iu(\beta\hbar-\tau))/\hbar} \eu{-\iu\op{H}_1(t-\iu\tau)/\hbar} \right],
	\label{Cff}
\end{align}
$Z_0=\Tr[\eu{-\beta\op{H}_0}]$ is the reactant partition function
and $\tau$ can in principle be any real number
but is typically chosen in the range $[0,\beta\hbar]$ for numerical stability.
The same formulation can also be obtained from linear-response theory \cite{ChandlerGreen,nonoscillatory}.

Expanding the trace in a coordinate-space representation gives
\begin{multline}
	k Z_0
	= \frac{\Delta^2}{\hbar^2} \iiint_{-\infty}^\infty K_0(\mathsf{x}',\mathsf{x}'',-t-\iu(\beta\hbar-\tau))
	\\
	\times K_1(\mathsf{x}'',\mathsf{x}',t-\iu\tau) \, \rmd\mathsf{x}' \rmd\mathsf{x}'' \rmd t,
\end{multline}
where the quantum propagator is defined as
\begin{align}
	K_n(\mathsf{x}',\mathsf{x}'',t) = \braket{\mathsf{x}' | \eu{-\iu\op{H}_n t/\hbar} | \mathsf{x}''}.
\end{align}

In order to derive the instanton approximation to this rate constant, we replace the exact quantum propagators by van-Vleck semiclassical propagators,
and employing steepest-descent integration over all three dummy variables.
This is most easily done by choosing a value of $\tau$ such that the stationary point is at $t=0$.

With an imaginary time argument, the van-Vleck propagator is given by the approximation \cite{Miller1971density,GutzwillerBook}
\begin{align}
	K_n(\mathsf{x}',\mathsf{x}'',-\iu\tau_n)
	\sim \sqrt\frac{C_n}{(2\pi\hbar)^f} \, \eu{-S_n/\hbar},
\end{align}
where $S_n$ is the Euclidean action \cite{Kleinert} along the classical trajectory from $\mathsf{x}(0)=\mathsf{x}'$ to $\mathsf{x}(\tau_n)=\mathsf{x}''$ in imaginary-time $\tau_n$,
\begin{align}
	S_n \equiv S_n(\mathsf{x}',\mathsf{x}'',\tau_n) = \int_0^{\tau_n} \left[ \half m \left(\pder{x}{\tau'}\right)^2 + V(x(\tau')) \right] \rmd \tau',
\end{align}
and the prefactor is given by
\begin{align}
	C_n = \left|-\frac{\partial^2 S_n}{\partial \mathsf{x}' \partial \mathsf{x}''}\right|.
\end{align}
This approximation is equivalent to taking a steepest-descent integration of all the beads in a discretized path-integral representation of the imaginary-time propagator.

Stationary points are given by $\pder{S}{\mathsf{x}'}=\pder{S}{\mathsf{x}''}=\pder{S}{\tau}=0$,
where $S=S_0+S_1$ is the sum of the actions of the two imaginary-time trajectories.
Thus the two trajectories together form a periodic orbit
in the classically forbidden region under the barrier,
which resembles the original instanton formulation
\cite{Miller1975semiclassical}.
The optimal hopping point, $\mathsf{x}^\ddag=\mathsf{x}'=\mathsf{x}''$, is defined as the intersection of the trajectories,
and obeys $V_0(\mathsf{x}^\ddag)=V_1(\mathsf{x}^\ddag)$.

In this way, the formula for the instanton rate constant is obtained as
\begin{align}
k_\text{inst} Z_0 = \sqrt{2 \pi \hbar} \, \frac{\Delta^2}{\hbar^2} \sqrt{\frac{C_0 C_1}{- \Sigma}} \, \eu{-S/\hbar},
\label{instanton}
\end{align}
where
\begin{align}
\Sigma = 
\begin{vmatrix}
\frac{\partial^2 S}{\partial \mathsf{x}' \partial \mathsf{x}'} & \frac{\partial^2 S}{\partial \mathsf{x}' \partial \mathsf{x}''} & \frac{\partial^2 S}{\partial \mathsf{x}' \partial \tau} \\
\frac{\partial^2 S}{\partial \mathsf{x}'' \partial \mathsf{x}'} & \frac{\partial^2 S}{\partial \mathsf{x}'' \partial \mathsf{x}''} & \frac{\partial^2 S}{\partial \mathsf{x}'' \partial \tau} \\
\frac{\partial^2 S}{\partial \tau \partial \mathsf{x}'} & \frac{\partial^2 S}{\partial \tau \partial \mathsf{x}''} & \frac{\partial^2 S}{\partial \tau^2}
\end{vmatrix}.
\end{align}
An equivalent steepest-descent approximation to the reactant partition function gives
\begin{equation}
	Z_0 = \prod_{k=1}^f \left[ 2 \sinh \frac{\beta \hbar \omega_k}{2} \right]^{-1},
\end{equation}
where $\omega_k$ are the normal mode frequencies at the minimum of $V_0(\mathsf{x})$.
The final expression we obtain for the rate is identical to that derived in \Ref{GoldenGreens} and similar to that derived in a different way in \Ref{Cao1997nonadiabatic}.

Note that here we have taken steepest-descent approximations for the nuclear coordinates and the time variables simultaneously,
and therefore the instanton approach in general gives a different result from that suggested by Wolynes \cite{Wolynes1987nonadiabatic}
which is a form of quantum instanton approach \cite{Miller2003QI,Vanicek2005QI}
and performs a steepest-descent integration in time only and obtains the nuclear fluctuations from path-integral Monte Carlo sampling.
For the particular system-bath model which we study in this paper, the reactant and product Hamiltonians, $\op{H}_n$, are quadratic
such that a steepest-descent integration over the coordinates is exact
and the rates obtained instanton expression will be equivalent to those from the method of Wolynes. \cite{GoldenGreens,GoldenRPI}

As in the standard ring-polymer instanton approaches \cite{Andersson2009Hmethane,RPinst,Althorpe2011ImF,AdiabaticGreens,HCH4,tunnel,water,octamer,hexamerprism,Kaestner2014review},
in order to obtain the instanton trajectory numerically,
we applied a ring-polymer discretization to the path-integral.
Equal imaginary-time intervals were used according to the Lagrangian formalism described in detail in \Ref{GoldenRPI}.
The nuclear configurations and the value of $\tau$ were optimized simultaneously using a saddle-point finding algorithm \cite{Nichols1990mep} in the space of $\{\mathbf{x},\tau\}$.
Numerical algorithms for computing the partial derivatives from discretized instanton trajectories are given explicitly in the Appendix of \Ref{GoldenRPI}. \cite{Kleinert,Althorpe2011ImF}

Unlike for the case of instantons on a single Born-Oppenheimer surface,
here there is no cross-over temperature 
\cite{Benderskii}
and so the approach is valid for all temperatures
and does not require corrections to match with the correct high-temperature limit.
\cite{Haenggi1991instanton,Kryvohuz2011rate,faraday}

\subsection{Classical limit}
\label{sec:cl_inst}
The classical limit of the instanton rate can be found
\cite{GoldenGreens}
in the limit of high temperature ($\beta \rightarrow 0$) where the instanton shrinks to an infinitesimally small line located at the minimum of the crossing seam.
The exponent is thus equal to $\beta V^\ddagger$ where $V^\ddagger$ is the activation energy as would be expected
from classical transition-state theory arguments.

Approximating the potentials in a Taylor series around this point, $\mathsf{x}^\ddagger$, gives
\begin{multline}
	V_n(\mathsf{x}) \approx V^{\ddagger} + \mathsf{g}_n^\T (\mathsf{x} - \mathsf{x}^{\ddagger})
	+ \thalf m (\mathsf{x} - \mathsf{x}^{\ddagger})^\T \mathsf{A}_n (\mathsf{x} - \mathsf{x}^{\ddagger})
\end{multline}
and the rate formula generalizes to
\begin{subequations}
\label{classical}
\begin{align}
	k_\text{cl} &= \sqrt{\frac{2 \pi m}{\beta \hbar^2}} \frac{\Delta^2}{\hbar |g_0 - g_1|} \frac{Z^{\ddagger}}{Z_0^\text{cl}} \, \eu{-\beta V^{\ddagger}}
	\\
	Z^{\ddagger} &= \det{}' \left[ \beta^2 \hbar^2
	\frac{g_0 \mathsf{A}_1 - g_1 \mathsf{A}_0}{g_0 - g_1}
	\right]^{-1/2}
	\label{Zddag}
	\\
	Z_0^\text{cl} &= \det \left[ \beta^2\hbar^2\mathsf{A}_0 \right]^{-1/2},
\end{align}
\end{subequations}
where $g_n=|\mathsf{g}_n|$ and the determinant in \eqn{Zddag} is taken after the reaction coordinate is projected out.
The reaction coordinate is
defined in this case to be parallel to the vector $\mathsf{g}_0$ or equivalently to $\mathsf{g}_1$ at the transition state.
The procedure is defined in more detail in \Ref{GoldenGreens}.
Note that this reaction coordinate is in the same direction as the energy gap coordinate,
at least at the transition state,
although we did not have to assume this to be true
but found it to be so automatically from the derivation.

$V^{\ddagger}$ is the value of the potential at the minimum of the crossing seam, \textit{i.e.}\ the potential energy of the classical transition state, $\mathsf{x}^\ddag$.
This value has to be found numerically in the general case.
For an $f$-dimensional system it is the minimum of the $(f-1)$-dimensional crossing seam defined by $V_0(\mathsf{x}) = V_1(\mathsf{x})$.

As noted in \Ref{GoldenGreens} this rate is a steepest descent version of a more general classical rate
derived from the classical limit of the flux correlation function formalism \cite{nonoscillatory}
but also discussed in older literature \cite{UlstrupBook}.
It has the simple form of the classical transition-state theory rate constant multiplied by twice the Landau-Zener hopping probability. \cite{Zener1932LZ,Nitzan}

In the case that the free-energy surfaces are harmonic, the classical rate constant reduces to Marcus theory exactly.
However, in general it is not possible to reformulate it in a simple form depending only on reorganization energies without making further approximations.
Instead, as in the Arrhenius equation, it is the activation energy (or activation free-energy) which is the dominant contributing variable,
which can only be computed rigorously from a molecular dynamics simulation
constrained to the crossing seam \cite{Hwang1987ET,Kuharski1988Fe3e,King1990ET}.

\section{Application to an asymmetric system-bath model}

Here we apply the theories discussed in the previous section to a simple model for electron transfer which 
exhibits anharmonic free-energy curves.

\subsection{Definition of the Model}

The potential-energy surfaces which appear in the Hamiltonians \eqn{Hamiltonian} for the asymmetric system-bath model are defined as
\begin{align}
V_n(\mathsf{x}) = V^\text{s}_n(x_1) + V^\text{b}(x_1,\dots,x_f),
\end{align}
where
\begin{align}
V^\text{s}_0(x_1) &= \thalf m \Omega_0^2 (x_1 + \xi)^2 \\
V^\text{s}_1(x_1) &= \thalf m \Omega_1^2 (x_1 - \xi)^2 - \epsilon
\end{align}
and the bath including coupling to the system coordinate, $x_1$, is \cite{Caldeira1983dissipation} 
\begin{align}
	V^\text{b}(x_1,\dots,x_f) = \sum_{j= 2}^{f} \thalf m \omega_j^2 \left( x_j - \frac{c_j}{m\omega_j^2} x_1 \right)^2.
\end{align}

The bath is defined in terms of its spectral density, \cite{Weiss}
\begin{equation}
J(\omega) = \frac{\pi}{2} \sum_{j = 2}^{f} \frac{c_j^2}{m \omega_j} \delta(\omega - \omega_j),
\end{equation}
where \cite{RPMDrate}
\begin{subequations}
\begin{align}
	\omega_j &= - \omega_c \log\left[(j-3/2)/(f-1)\right]
	\\
	c_j &= m \omega_j \sqrt{2 \gamma \omega_c /\pi (f-1)}
\end{align}
\end{subequations}
for $j=2\dots f$.
In the continuum limit, when $f\rightarrow\infty$, %
this discretization scheme tends to an Ohmic spectral density, \cite{Weiss}
\begin{equation}
	J(\omega) = m \omega \gamma \eu{-\omega/\omega_c}.
\end{equation}
Here $\omega_c$ is the cut-off frequency and $\gamma$
the friction coefficient.

In many previous studies,
ET has been described by a spin-boson model, where all modes are coupled linearly to the reactant and product states.
The relationship between system-bath models and the more common spin-boson description of ET
is described in detail in \Ref{Thoss2001hybrid}.
For the symmetric case where $\Omega_0 = \Omega_1$, it can be shown that this system-bath Hamiltonian is equivalent to that of the usual spin-boson model.
However, they cannot be mapped onto each other in general.

Note that in the usual Marcus theory, the term asymmetric refers to cases where $\epsilon \ne 0$.
Here, we use the term asymmetric differently, such that it instead refers to the case where the reactant and product reorganization energies are not equal, \textit{i.e.}\ $\lambda_0 \ne \lambda_1$,
regardless of whether or not $\epsilon$ is 0.
As shall be shown, the two reorganization energies are only equivalent in the symmetric case.

\subsection{Free-energy curves}

The free-energy curves for an ET system can be obtained from the probability distributions of the energy gap fluctuations along the reaction coordinate, $\mathcal{E}$, \cite{ChandlerET}
\begin{align}
	P_n(\mathcal{E}) &= \Braket{ \delta\left(\mathcal{E} - \thalf \left[V_0(\mathsf{x}) - V_1(\mathsf{x})\right] \right) }_n
	\\
	&= \frac{\int \delta\left(\mathcal{E} - \frac{1}{2}\left[V_0(\mathsf{x}) - V_1(\mathsf{x})\right] \right) \eu{-\beta V_n} \, \rmd\mathsf{x} }{ \int \eu{-\beta V_n} \, \rmd\mathsf{x} }.
\end{align}
The free-energy curves are then given by
\begin{equation}
	F_n(\mathcal{E}) = - \frac{1}{\beta} \ln (P_n(\mathcal{E})).
\end{equation}

\begin{figure}
\includegraphics{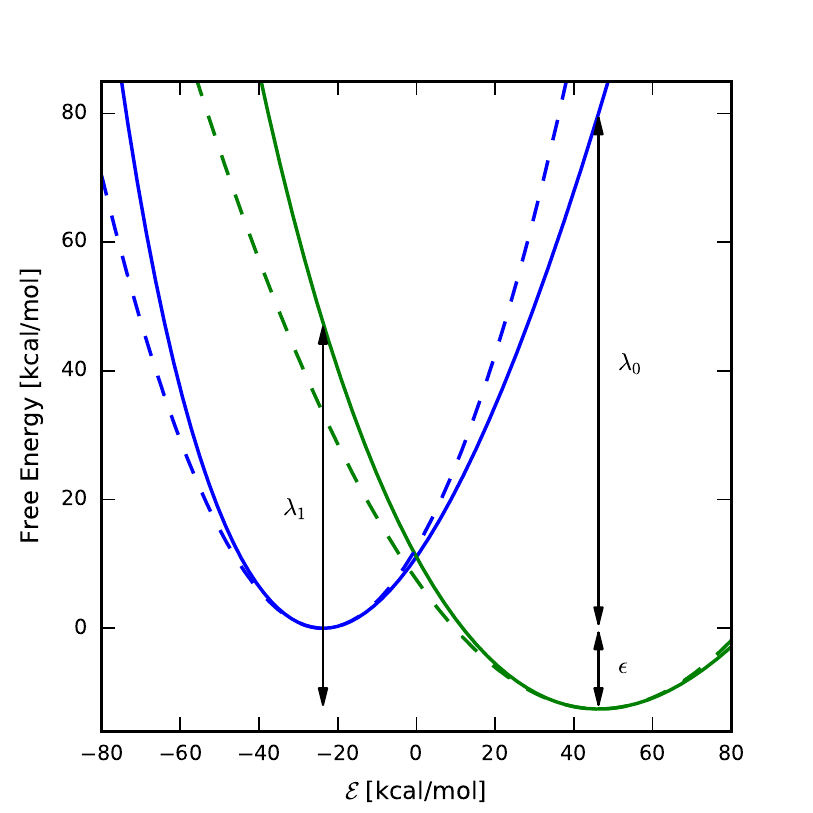}
\caption{Free-energy curves along the energy-gap reaction coordinate, $\mathcal{E}$, for an asymmetric system.
The reactant is depicted in blue on the left and the product in green on the right.
Dashed lines show a harmonic approximation about the minimum of each free-energy curve, indicating that the curves are slightly anharmonic.
The definitions of the reorganization energies and bias (all positive values) are indicated by arrows.
}
\label{fig:curves}
\end{figure}

For our system, within a small logarithmic correction, the free-energy curves are simply given by the system potential, $F_n(\mathcal{E}) = V^\text{s}_n(x_1(\mathcal{E}))$,
where $x_1(\mathcal{E})$ solves %
\begin{align}
	\mathcal{E} = \thalf\left[ V_0^\text{s}(x_1(\mathcal{E})) - V_1^\text{s}(x_1(\mathcal{E})) \right].
\end{align}
If two or more solutions exist, the solution with the lowest energy is taken.
Although $V^\text{s}_n(x_1)$ are harmonic, for an asymmetric system the free-energy curves plotted with respect to $\mathcal{E}$ are not.
\Fig{curves} shows the shape of the free-energy curves for an example system.
From the free-energy curves the reorganization energies $\lambda_0$ and $\lambda_1$ can be obtained as shown by the arrows.

For this system, for which the free-energy curves are known analytically,
the minima of the curves are given by $x_1=-\xi$ and $x_1=\xi$, or equivalently $\mathcal{E}=\epsilon/2-m\Omega_1^2\xi^2$ and $\mathcal{E}=\epsilon/2+m\Omega_0^2\xi^2$, such that the reorganization energies are
\begin{align}
	\label{lambda}
	\lambda_n = 2 m \Omega_n^2 \xi^2.
\end{align}

For the symmetric Marcus case, when $\lambda_0$ and $\lambda_1$ are equal, both curves become parabolas with the same frequency.
For asymmetric systems however, two different reorganization energies, $\lambda_0$ and $\lambda_1$, are found.  \cite{Schwerdtfeger2014ET}
Situations like this have been found in a number of computer simulations of %
molecular systems.
\cite{Tachiya1989ET,Blumberger2006Ag,Krapf2012Marcus}
The standard Marcus theory formula, \eqn{Marcus}, should not be applied in this case and we should use a more general formulation instead such as instanton theory,
or if quantum effects can be neglected, the classical limit of instanton theory.

In the standard Marcus model,
there are alternative methods for obtaining the reorganization energies
based on statistics of the average and fluctuations of the energy gap.
\cite{ChandlerET,Blumberger2006RuRu}
For this system-bath model 
it would also be possible to obtain the values for the two reorganization energies and the bias
only from equilibrium simulations of the reactant or product.
Although the free-energy curves are not harmonic, they are approximately so near their equilibria. \cite{Tachiya1989ET}
Thus it is necessary for only a minor extension to the standard theory to take account of the fact
that the curvature is different in the reactant and product case.

We define the instantaneous energy gap as $\Delta V(\mathsf{x}) = V_0(\mathsf{x}) - V_1(\mathsf{x})$.
The mean of this variable is
\begin{subequations}
\label{mean}
\begin{align}
	\braket{\Delta V}_0 &= \epsilon - \lambda_1 + O(\beta^{-1})
	\\
	\braket{\Delta V}_1 &= \epsilon + \lambda_0 + O(\beta^{-1})
\end{align}
\end{subequations}
and its standard deviation is
\begin{subequations}
\begin{align}
	\sigma_0^2 &= \braket{(\Delta V - \braket{\Delta V}_0)^2}_0 = \frac{2\lambda_1^2}{\beta\lambda_0} + O(\beta^{-2})
	\\
	\sigma_1^2 &= \braket{(\Delta V - \braket{\Delta V}_1)^2}_1 = \frac{2\lambda_0^2}{\beta\lambda_1} + O(\beta^{-2}).
\end{align}
\end{subequations}
By neglecting terms with higher orders of $\beta^{-1}$, we are assuming that the system is at a low enough temperature
that the harmonic approximation around each of the free-energy curves is valid within the energy range sampled by the equilibrium distribution.
This is an excellent approximation at room temperature (for which $\beta^{-1}\approx 0.6$ kcal/mol) as can be seen from \fig{curves}.
In the symmetric theory, the reorganization energies can thus be defined as \cite{Blumberger2006RuRu}
$\beta \sigma_n^2/2$,
but for asymmetric systems this is clearly not equivalent to $\lambda_n$.
By solving the simultaneous equations,
the correct reorganization energies can however be recovered as
\begin{subequations}
\label{lam}
\begin{align}
	\lambda_0 &= \thalf \beta \sigma_0^{2/3} \sigma_1^{4/3}
	\\
	\lambda_1 &= \thalf \beta \sigma_0^{4/3} \sigma_1^{2/3}
\end{align}
\end{subequations}
and hence also the product bias $\epsilon$ from \eqn{mean}.

Note that the equivalence of \eqn{lam} and \eqn{lambda} only formally holds
for our system-bath model.
Nonetheless, for more complex systems it may be a good approximation
and this approach may be useful
when the free-energy surfaces are not known \textit{a priori}.
It will in any case be more accurate than the standard approach
which assumes a symmetric form.

\subsection{Classical transition-state theory rate}

The classical transition-state theory rate constant, \eqn{classical},
can be evaluated analytically for this system
to give a definition in terms of only the reorganization energies and bias.

First we define an asymmetry parameter
\begin{align}
	\alpha = \frac{\lambda_0-\lambda_1}{\lambda_0+\lambda_1},
\end{align}
which varies between $-1$ and $1$ and is $0$ for the symmetric case.
The transition state, defined as the minimum of the crossing seam, is located at
\begin{align}
	x_1^\ddag/\xi = -\frac{1}{\alpha} + \frac{1}{\alpha} \sqrt{1 - \alpha\left(\alpha + \frac{4\epsilon}{\lambda_0+\lambda_1}\right)}
\end{align}
with all other modes $x_j^\ddag=x_1^\ddag$ for $j\ge2$.
This gives the following formulas for the activation energy,
\begin{align}
	V^\ddag = V_0^\text{s}(x_1^\ddag) = V_1^\text{s}(x_1^\ddag)
	= \frac{1}{4} \lambda_0 (x_1^\ddag/\xi+1)^2,
\end{align}
and gradients,
\begin{align}
	g_0 &= \sqrt\frac{m\lambda_0}{2}\Omega_0(x_1^\ddag/\xi+1)
	\\
	g_1 %
		&= \sqrt\frac{m\lambda_0}{2}\Omega_0\frac{1-\alpha}{1+\alpha}(x_1^\ddag/\xi-1).
\end{align}
Finally, the ratio of partition function is $Z^\ddag/Z_0=\beta\hbar\Omega_0$,
such that we obtain the expression
\begin{align} \label{classical2}
	k_\text{cl} = \frac{\Delta^2}{\hbar} \sqrt\frac{\pi\beta}{\lambda_0} \frac{1+\alpha}{1+\alpha x_1^\ddag/\xi} \eu{-\beta V^\ddag}.
\end{align}
In the limit of $\alpha\rightarrow0$, this reduces to the usual Marcus theory rate constant, \eqn{Marcus}. %

This formula provides a simple asymmetric generalization of Marcus theory,
similar to the approach taken in \Ref{Krapf2012Marcus}.
It assumes that the reactant and product can be effectively described by harmonic oscillators of differing frequencies.
Armed only with values for both reorganization energies and the bias obtained from equilibrium simulations as described in Eqs.~(\ref{mean}--\ref{lam}), this formula could also be applied to more general systems.

Here there is no effect from friction of the bath.
This is because the transition-state theory is not of the usual kind which measures flux through a dividing surface in the system coordinate
\cite{Wigner1938TST}
and would therefore
feel friction from the other modes \cite{Pollak1986Kramers}.
Instead this rate measures the flux from one electronic state to another \cite{nonoscillatory} and is therefore not affected by the bath friction,
at least within the classical approximation.

\subsection{Results}

In our calculations, we investigate the effect of asymmetry on the rates.
To this end, the value of $\lambda_0$ is kept constant while the $\lambda_1$ is changed.
As is commonly done in ET studies, the standard Marcus theory rate constant, \eqn{Marcus}, is defined using $\lambda=\lambda_0$
and is therefore not affected at all by the asymmetry.
However, we also compare with another simple approximation \cite{Muegge1997ET} using the same formula but with $\lambda=\lambda_\text{av}\equiv(\lambda_0+\lambda_1)/2$. 

The fixed parameters we choose
are given in \tref{params}
and $\lambda_1$ was varied between 40 and 160 kcal/mol.
This defines a different value for $\Omega_1$ in each case but $\xi$ remains fixed to give the appropriate value of $\lambda_0$ according to \eqn{lambda}.
The mass parameter, $m$, does not affect any of the rates
and as all rates depend on $\Delta$ in the same way, this also does not affect our conclusions and so is not specified.
We simply assume that the coupling is small such that the golden-rule limit is reached.

\begin{table}
\caption{Fixed parameters used to define the system-bath model used in the calculations.  Various values for the product reorganization energy, $\lambda_1$ are chosen for study.
The remaining parameters of $m$ and $\Delta$ do not affect the results and are thus free parameters.}
\label{tab:params}
\begin{tabular}{cc}
\hline
Parameter & Value \\
\hline
$\epsilon$ & 0 or 40 kcal/mol \\
$\lambda_0$ & 80 kcal/mol \\
$\lambda_1$ & 40--160 kcal/mol \\
$\Omega_0$ & 500 cm$^{-1}$ \\
$\omega_c$ & 500 cm$^{-1}$ \\
$\gamma$ & $0.001$ a.u. \\
$f$ & 8 \\
$T$ & 300 K \\
\hline
\end{tabular}
\end{table}

These parameters have been chosen to replicate a 
typical electron-transfer process in solution.
We test two models, one without a bias, \textit{i.e.}\ $\epsilon=0$, and the other with a bias, $\epsilon=40\, \mathrm{kcal/mol}$
Although it is unlikely that a real system with different frequencies in the reactant and product states would not have a bias,
we test the unbiased system so that we can observe the asymmetric effect independently of the bias.
The behaviour with respect to bias is well known from the original Marcus theory such that the rate increases with bias in the normal regime.
Both the instanton and its classical limit can also be applied in exactly the same way for such systems.

\begin{figure}
\includegraphics{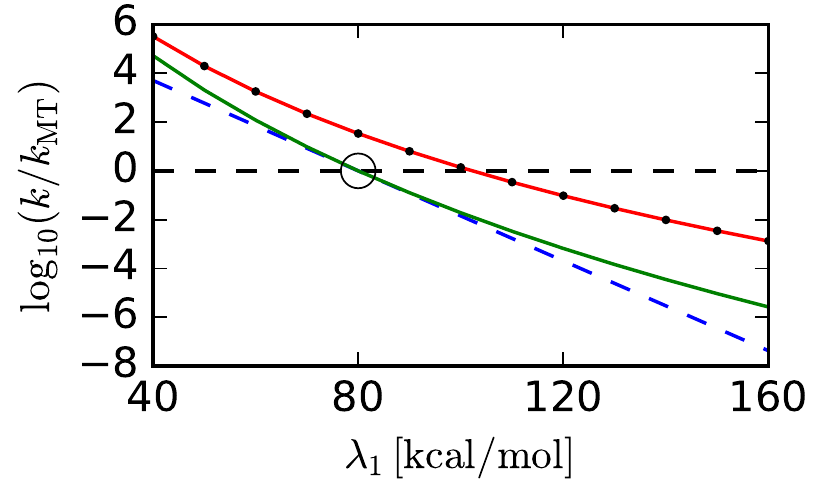}
\caption{Rate constants computed by the various methods for varying asymmetry on the unbiased system, $\epsilon=0$.
They are compared to the simplest Marcus theory approximation ($\lambda=\lambda_0$) which is in this case independent of $\lambda_1$ and is represented by the horizontal dashed black line.
The blue dashed line shows Marcus theory using the averaged reorganization energy $\lambda=\lambda_\text{av}$
and
the green solid line gives the classical rate constant from \eqn{classical}.
When $\lambda_1$ is equal to $\lambda_0$, all three classical theories agree and this point is highlighted by the open circle.
The red solid line follows the instanton rate constant from \eqn{instanton}
and the black dots give the exact rate constant computed from the method described in the Appendix.
}
\label{fig:rate}
\end{figure}

Alongside exact results,
the rate constants for the unbiased system
obtained by the four approximate methods are plotted in \fig{rate}
relative to the classical Marcus theory rate for the symmetric system.
As already discussed, the standard approach using $\lambda=\lambda_0$ is independent of $\lambda_1$.
The classical rate, however, which is our classical benchmark as it is rigorously derived to describe the asymmetry in this system,
while ignoring quantum nuclear effects,
shows a large variation with $\lambda_1$.
The standard Marcus approach can thus
give rate predictions which are incorrect by many orders of magnitude.

Somewhat surprisingly, Marcus theory with the averaged reorganization energy is seen to match fairly well to the classical results for weak asymmetry.
However, it is no more difficult to evaluate than \eqn{classical2} which should be used instead as it is more reliable for more asymmetric systems
where the ratio of reorganization energies is close to 2,
such as has been found in certain cases from simulations \cite{Blumberger2006Ag,Krapf2012Marcus}.

The instanton results extend the classical rate theory to show the quantum effects of the nuclear dynamics.
Due to tunnelling, the rate of the symmetric system is increased by a factor of 33.8.
This speed-up is even greater when $\lambda_1$ is increased
as this increases the value of $\Omega_1$
and makes the gradient near the transition-state steeper.
In turn this makes the width of the barrier region smaller which
makes tunnelling easier and increases the rate.

Unlike for the classical case, here the friction of the bath does affect the instanton rate.
As for the tunnelling problem in a double well, the friction inhibits the growth of the instanton
and slightly lessens the effect of tunnelling \cite{Caldeira1983dissipation,Wolynes1987dissipation,Kim2006bath}.
The speed-up due to tunnelling for the symmetric system was found to be 37.6 without friction.

We can explain the
reason for the seemingly good behaviour of Marcus theory with the averaged reorganization energy
in terms of a Taylor series of the activation energy on $\alpha$,
\begin{multline}
	V^\ddag \approx \frac{(\lambda_\text{av}-\epsilon)^2}{4\lambda_\text{av}}
		+ \frac{\epsilon(\epsilon^2 - \lambda_\text{av}^2)}{4\lambda_\text{av}^2} \alpha
		\\
		+ \frac{(\epsilon^2 + \lambda_\text{av}^2)(5\epsilon^2-3\lambda_\text{av}^2)}{16\lambda_\text{av}^3} \alpha^2
		+ O(\alpha^3).
\end{multline}
The first term with zeroth order in $\alpha$ is obviously the
same activation energy as is used in the Marcus theory with $\lambda=\lambda_\text{av}$.
For a system with no bias,
such as we have tested here,
the linear correction term is identically zero.
Thus only for relatively strong asymmetry is a deviation seen from the benchmark classical results.
However, this is not true of a system with a bias,
so we predict that for a biased system
the result will not be in such good agreement,
even for relatively weak asymmetry.
Asymmetric effects are most important near $\epsilon=\lambda_\text{av}/\sqrt{3}$ which maximizes the first-order term.
When the bias increases further and becomes approximately equal to the average reorganization energy,
which is the activationless regime in the symmetric case,
the asymmetry becomes a second-order effect as the linear term approaches zero.
For stronger biases again, one enters the
inverted regime and asymmetry acts in the opposite direction,
\textit{i.e.}\ decreasing the rate for $\lambda_0>\lambda_1$ and increasing it for $\lambda_0<\lambda_1$.
Note that this analysis is only valid for small values of $\alpha$ and outside these limits,
the effects of asymmetry can be large for any system.
Nonetheless the analysis is quite general and we expect the trends to also apply to realistic systems
without the explicit system-bath Hamiltonian treated here.

\begin{figure}
\includegraphics{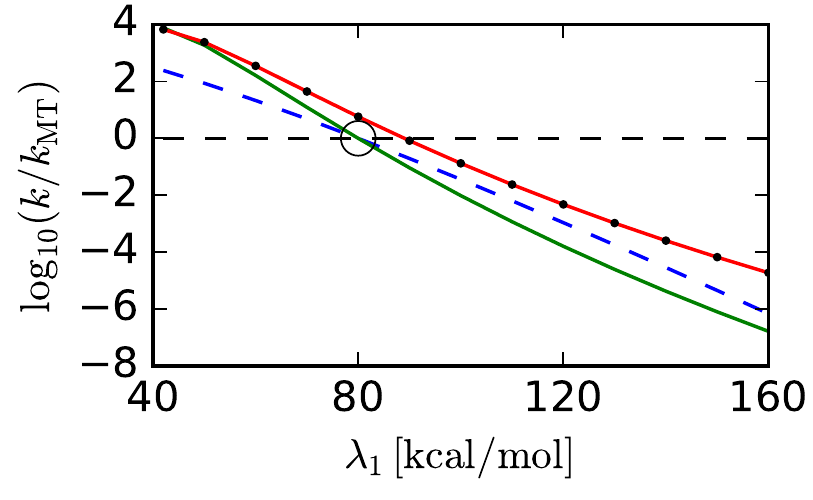}
\caption{As for \fig{rate} but for a biased system with $\epsilon=40\, \mathrm{kcal/mol}$.
}
\label{fig:biased}
\end{figure}

To confirm our predictions,
calculations were also performed for a biased system with $\epsilon=40\, \mathrm{kcal/mol}$.
As expected, and as shown in \fig{biased}, Marcus theory with the averaged reorganization energy
gives a poor description of the rate for all asymmetric systems.
The tunnelling effect is also reduced for this particular system compared with the unbiased system
and in particular is absent for $\lambda_1=40\,\mathrm{kcal/mol}$ which corresponds to the activationless regime.
Note that the relatively good agreement between Marcus theory with the averaged reorganization energy and the exact results for high $\lambda_1$ is fortuitous.
It comes from a cancellation of errors of neglecting tunnelling
as well as treating asymmetry approximately.

Comparison of the instanton results with numerically exact calculation is very encouraging.
It correctly describes the trend with asymmetry and also approximates the tunnelling factor accurately.
The error is less than 1\% for all systems studied and is equally
good at describing the asymmetric as the symmetric case.

All rate constants are converged with respect to the number of bath modes and increasing $f$ 
makes essentially no difference to the results.
The instanton results are also converged with respect to the number of ring-polymer beads.
We used 128 beads split over the two potentials in such a way as to give approximately the same spring constants in the ring-polymer potential.
Tests with larger numbers of beads again found no change to the first 3 significant figures of the rate constant.

\section{Conclusions}

The nonadiabatic instanton formulation has been found
to give an excellent description of the ET rate in an asymmetric system-bath model in the golden-rule limit.
In addition, the formula obtained from its classical limit is able to describe
the behaviour of the rate with asymmetry,
defined by the relative difference between the reactant and product reorganization energies.
The classical formula, \eqn{classical2}, provides a simple generalization of Marcus theory to treat asymmetric reactions %
and we show how all the necessary quantities can be obtained from equilibrium molecular dynamics simulations.

It can be expected that for more realistic systems with anharmonic potentials,
that the deviation of the nonadiabatic instanton prediction from exact results would increase slightly.
Typical errors caused by the standard instanton approximation are
seen to be around 20\% for hydrogen transfer in the gas-phase \cite{tunnel,formic,HCH4,faraday}
when exact calculations are available for comparison.
However, what is important is that the order of magnitude is consistently predicted correctly, \cite{water}
and trends are well described
such that, for instance, unknown mechanisms can be discovered. \cite{hexamerprism}

A good description of the tunnelling of the nuclei will be most
important for cases where
an individual proton rearrangement occurs simultaneously with the electron transfer. \cite{Tiwari2016Faraday}
The nonadiabatic instanton formulation would be a good candidate for studying
such proton-coupled electron transfer reactions
\cite{HammesSchiffer2015PCET}
in multidimensional systems.

The standard instanton approach can be used
to explain the successes \cite{RPinst,Hele2013QTST} and failures \cite{Shushkov2013instanton} of ring-polymer molecular dynamics \cite{Habershon2013RPMDreview}.
In the same way, a better understanding of the instanton approach for nonadiabatic transitions
should help the development of nonadiabatic ring-polymer molecular dynamics.
\cite{Shushkov2012RPSH,mapping,nonoscillatory,vibronic,Ananth2013MVRPMD,Hele2016Faraday,Duke2016Faraday,Chowdhury2017CSRPMD,Menzeleev2014kinetic}

\section{Acknowledgements}
The authors acknowledge the financial support from the Swiss National Science Foundation through the NCCR MUST (Molecular Ultrafast Science and Technology) Network. 
We would also like to thank David Reichman for
making us aware of the methods described in the Appendix used to calculate the exact rate.

\appendix
\section{Exact calculation of the rate}

Because our system is defined as a set of linearly-coupled harmonic oscillators, it is possible to compute the exact rate for the system.
In this case we use the same parameters and the same discretization of the bath into $f-1$ modes as used above.
Following the matrix method outlined in \Ref{Friesner1985Raman}
and using Eq.~(30) from \Ref{Balian1969Bogoliubov}
the flux correlation function, \eqn{Cff},
can be written
in the following form:
\begin{align}
	C^\tau(t) &= 
	\Tr\left[ \eu{\half \mathbf{\gamma}^\T \mathbf{S} \mathbf{\gamma} + \mathbf{\lambda}^\T \mathbf{\tau} \mathbf{\gamma} + \nu} \right]
	\label{Cff2}
	\\
	&= \left[ (-1)^f \det\left\{\eu{\mathbf{\tau} \mathbf{S}} - \mathbf{1}\right\} \right]^{-\half}
		\eu{\nu - \half\mathbf{\lambda}^\T \mathbf{\tau} \mathbf{S}^{-1} \mathbf{\tau}^{-1} \mathbf{\lambda}},
\end{align}
where $\mathbf{\gamma}=(\op{b}_1,\dots,\op{b}_f,\op{b}_1^\dagger,\dots,\op{b}_f^\dagger)$ is a $2f$-dimensional vector of boson creation and annihilation operators, 
and $\mathbf{\tau}$ is the following $2f\times2f$ rotation matrix:
\begin{align}
	\mathbf{\tau} = \begin{pmatrix} \mathbf{0} & \mathbf{1} \\ -\mathbf{1} & \mathbf{0} \end{pmatrix}.
\end{align}
The remaining three variables,
$\mathbf{S}$, a $2f\times2f$ symmetric matrix, $\mathbf{\lambda}$, a $2f$-dimensional vector, and
$\nu$, a scalar,
are uniquely defined by the parameters of the Hamiltonian.

Then performing the integral over $t$ in \eqn{exact} by quadrature gives the exact rate constant, $k$.
Again the value of $\tau$ used in the correlation function, \eqn{Cff}, can be any real number.
A typical choice is $\tau=\beta\hbar/2$.

\clubpenalty10000 %

\clearpage %


\begin{thebibliography}{99}%
\makeatletter
\providecommand \@ifxundefined [1]{%
 \@ifx{#1\undefined}
}%
\providecommand \@ifnum [1]{%
 \ifnum #1\expandafter \@firstoftwo
 \else \expandafter \@secondoftwo
 \fi
}%
\providecommand \@ifx [1]{%
 \ifx #1\expandafter \@firstoftwo
 \else \expandafter \@secondoftwo
 \fi
}%
\providecommand \natexlab [1]{#1}%
\providecommand \enquote  [1]{``#1''}%
\providecommand \bibnamefont  [1]{#1}%
\providecommand \bibfnamefont [1]{#1}%
\providecommand \citenamefont [1]{#1}%
\providecommand \href@noop [0]{\@secondoftwo}%
\providecommand \href [0]{\begingroup \@sanitize@url \@href}%
\providecommand \@href[1]{\@@startlink{#1}\@@href}%
\providecommand \@@href[1]{\endgroup#1\@@endlink}%
\providecommand \@sanitize@url [0]{\catcode `\\12\catcode `\$12\catcode
  `\&12\catcode `\#12\catcode `\^12\catcode `\_12\catcode `\%12\relax}%
\providecommand \@@startlink[1]{}%
\providecommand \@@endlink[0]{}%
\providecommand \url  [0]{\begingroup\@sanitize@url \@url }%
\providecommand \@url [1]{\endgroup\@href {#1}{\urlprefix }}%
\providecommand \urlprefix  [0]{URL }%
\providecommand \Eprint [0]{\href }%
\providecommand \doibase [0]{http://dx.doi.org/}%
\providecommand \selectlanguage [0]{\@gobble}%
\providecommand \bibinfo  [0]{\@secondoftwo}%
\providecommand \bibfield  [0]{\@secondoftwo}%
\providecommand \translation [1]{[#1]}%
\providecommand \BibitemOpen [0]{}%
\providecommand \bibitemStop [0]{}%
\providecommand \bibitemNoStop [0]{.\EOS\space}%
\providecommand \EOS [0]{\spacefactor3000\relax}%
\providecommand \BibitemShut  [1]{\csname bibitem#1\endcsname}%
\let\auto@bib@innerbib\@empty
\bibitem [{\citenamefont {Marcus}(1993)}]{Marcus1993review}%
  \BibitemOpen
  \bibfield  {author} {\bibinfo {author} {\bibfnamefont {R.~A.}\ \bibnamefont
  {Marcus}},\ }\href {\doibase 10.1103/RevModPhys.65.599} {\bibfield  {journal}
  {\bibinfo  {journal} {Rev. Mod. Phys.}\ }\textbf {\bibinfo {volume} {65}},\
  \bibinfo {pages} {599} (\bibinfo {year} {1993})}\BibitemShut {NoStop}%
\bibitem [{\citenamefont {Kuharski}\ \emph {et~al.}(1988)\citenamefont
  {Kuharski}, \citenamefont {Bader}, \citenamefont {Chandler}, \citenamefont
  {Sprik}, \citenamefont {Klein},\ and\ \citenamefont
  {Impey}}]{Kuharski1988Fe3e}%
  \BibitemOpen
  \bibfield  {author} {\bibinfo {author} {\bibfnamefont {R.~A.}\ \bibnamefont
  {Kuharski}}, \bibinfo {author} {\bibfnamefont {J.~S.}\ \bibnamefont {Bader}},
  \bibinfo {author} {\bibfnamefont {D.}~\bibnamefont {Chandler}}, \bibinfo
  {author} {\bibfnamefont {M.}~\bibnamefont {Sprik}}, \bibinfo {author}
  {\bibfnamefont {M.~L.}\ \bibnamefont {Klein}}, \ and\ \bibinfo {author}
  {\bibfnamefont {R.~W.}\ \bibnamefont {Impey}},\ }\href {\doibase
  10.1063/1.454929} {\bibfield  {journal} {\bibinfo  {journal} {J.~Chem.
  Phys.}\ }\textbf {\bibinfo {volume} {89}},\ \bibinfo {pages} {3248} (\bibinfo
  {year} {1988})}\BibitemShut {NoStop}%
\bibitem [{\citenamefont {Blumberger}(2008)}]{Blumberger2008ET}%
  \BibitemOpen
  \bibfield  {author} {\bibinfo {author} {\bibfnamefont {J.}~\bibnamefont
  {Blumberger}},\ }\href {\doibase 10.1039/b807444e} {\bibfield  {journal}
  {\bibinfo  {journal} {Phys. Chem. Chem. Phys.}\ }\textbf {\bibinfo {volume}
  {10}},\ \bibinfo {pages} {5651} (\bibinfo {year} {2008})}\BibitemShut
  {NoStop}%
\bibitem [{\citenamefont {Chandler}(1998)}]{ChandlerET}%
  \BibitemOpen
  \bibfield  {author} {\bibinfo {author} {\bibfnamefont {D.}~\bibnamefont
  {Chandler}},\ }in\ \href@noop {} {\emph {\bibinfo {booktitle} {Classical and
  Quantum Dynamics in Condensed Phase Simulations}}},\ \bibinfo {editor}
  {edited by\ \bibinfo {editor} {\bibfnamefont {B.~J.}\ \bibnamefont {Berne}},
  \bibinfo {editor} {\bibfnamefont {G.}~\bibnamefont {Ciccotti}}, \ and\
  \bibinfo {editor} {\bibfnamefont {D.~F.}\ \bibnamefont {Coker}}}\ (\bibinfo
  {publisher} {World Scientific},\ \bibinfo {address} {Singapore},\ \bibinfo
  {year} {1998})\ Chap.~\bibinfo {chapter} {2}, pp.\ \bibinfo {pages}
  {25--49}\BibitemShut {NoStop}%
\bibitem [{\citenamefont {Zwanzig}(2001)}]{Zwanzig}%
  \BibitemOpen
  \bibfield  {author} {\bibinfo {author} {\bibfnamefont {R.}~\bibnamefont
  {Zwanzig}},\ }\href@noop {} {\emph {\bibinfo {title} {Nonequilibrium
  Statistical Mechanics}}}\ (\bibinfo  {publisher} {Oxford University Press},\
  \bibinfo {year} {2001})\BibitemShut {NoStop}%
\bibitem [{\citenamefont {Marcus}(1956)}]{Marcus1956ET}%
  \BibitemOpen
  \bibfield  {author} {\bibinfo {author} {\bibfnamefont {R.~A.}\ \bibnamefont
  {Marcus}},\ }\href {\doibase 10.1063/1.1742723} {\bibfield  {journal}
  {\bibinfo  {journal} {J.~Chem. Phys.}\ }\textbf {\bibinfo {volume} {24}},\
  \bibinfo {pages} {966} (\bibinfo {year} {1956})}\BibitemShut {NoStop}%
\bibitem [{\citenamefont {Marcus}(1960)}]{Marcus1960ET}%
  \BibitemOpen
  \bibfield  {author} {\bibinfo {author} {\bibfnamefont {R.~A.}\ \bibnamefont
  {Marcus}},\ }\href {\doibase 10.1039/DF9602900021} {\bibfield  {journal}
  {\bibinfo  {journal} {Discuss. Faraday Soc.}\ }\textbf {\bibinfo {volume}
  {29}},\ \bibinfo {pages} {21} (\bibinfo {year} {1960})}\BibitemShut {NoStop}%
\bibitem [{\citenamefont {Marcus}(1964)}]{Marcus1964review}%
  \BibitemOpen
  \bibfield  {author} {\bibinfo {author} {\bibfnamefont {R.~A.}\ \bibnamefont
  {Marcus}},\ }\href@noop {} {\bibfield  {journal} {\bibinfo  {journal} {Annu.
  Rev. Phys. Chem.}\ }\textbf {\bibinfo {volume} {15}},\ \bibinfo {pages} {155}
  (\bibinfo {year} {1964})}\BibitemShut {NoStop}%
\bibitem [{\citenamefont {Marcus}\ and\ \citenamefont
  {Sutin}(1985)}]{Marcus1985review}%
  \BibitemOpen
  \bibfield  {author} {\bibinfo {author} {\bibfnamefont {R.~A.}\ \bibnamefont
  {Marcus}}\ and\ \bibinfo {author} {\bibfnamefont {N.}~\bibnamefont {Sutin}},\
  }\href {\doibase 10.1016/0304-4173(85)90014-X} {\bibfield  {journal}
  {\bibinfo  {journal} {Biochim. Biophys. Acta}\ }\textbf {\bibinfo {volume}
  {811}},\ \bibinfo {pages} {265} (\bibinfo {year} {1985})}\BibitemShut
  {NoStop}%
\bibitem [{\citenamefont {Tachiya}(1989)}]{Tachiya1989ET}%
  \BibitemOpen
  \bibfield  {author} {\bibinfo {author} {\bibfnamefont {M.}~\bibnamefont
  {Tachiya}},\ }\href@noop {} {\bibfield  {journal} {\bibinfo  {journal}
  {J.~Phys. Chem.}\ }\textbf {\bibinfo {volume} {93}},\ \bibinfo {pages} {7050}
  (\bibinfo {year} {1989})}\BibitemShut {NoStop}%
\bibitem [{\citenamefont {Miller}, \citenamefont {Calcaterra},\ and\
  \citenamefont {Closs}(1984)}]{Miller1984inverted}%
  \BibitemOpen
  \bibfield  {author} {\bibinfo {author} {\bibfnamefont {J.~R.}\ \bibnamefont
  {Miller}}, \bibinfo {author} {\bibfnamefont {L.~T.}\ \bibnamefont
  {Calcaterra}}, \ and\ \bibinfo {author} {\bibfnamefont {G.~L.}\ \bibnamefont
  {Closs}},\ }\href {\doibase 10.1021/ja00322a058} {\bibfield  {journal}
  {\bibinfo  {journal} {J.~Am. Chem. Soc.}\ }\textbf {\bibinfo {volume}
  {106}},\ \bibinfo {pages} {3047} (\bibinfo {year} {1984})}\BibitemShut
  {NoStop}%
\bibitem [{\citenamefont {Hwang}\ and\ \citenamefont
  {Warshel}(1987)}]{Hwang1987ET}%
  \BibitemOpen
  \bibfield  {author} {\bibinfo {author} {\bibfnamefont {J.~K.}\ \bibnamefont
  {Hwang}}\ and\ \bibinfo {author} {\bibfnamefont {A.}~\bibnamefont
  {Warshel}},\ }\href@noop {} {\bibfield  {journal} {\bibinfo  {journal} {J.
  Am. Chem. Soc.}\ }\textbf {\bibinfo {volume} {109}},\ \bibinfo {pages} {715}
  (\bibinfo {year} {1987})}\BibitemShut {NoStop}%
\bibitem [{\citenamefont {King}\ and\ \citenamefont
  {Warshel}(1990)}]{King1990ET}%
  \BibitemOpen
  \bibfield  {author} {\bibinfo {author} {\bibfnamefont {G.}~\bibnamefont
  {King}}\ and\ \bibinfo {author} {\bibfnamefont {A.}~\bibnamefont {Warshel}},\
  }\href {\doibase 10.1063/1.459255} {\bibfield  {journal} {\bibinfo  {journal}
  {J.~Chem. Phys.}\ }\textbf {\bibinfo {volume} {93}},\ \bibinfo {pages} {8682}
  (\bibinfo {year} {1990})}\BibitemShut {NoStop}%
\bibitem [{\citenamefont {Blumberger}\ \emph {et~al.}(2006)\citenamefont
  {Blumberger}, \citenamefont {Tavernelli}, \citenamefont {Klein},\ and\
  \citenamefont {Sprik}}]{Blumberger2006Ag}%
  \BibitemOpen
  \bibfield  {author} {\bibinfo {author} {\bibfnamefont {J.}~\bibnamefont
  {Blumberger}}, \bibinfo {author} {\bibfnamefont {I.}~\bibnamefont
  {Tavernelli}}, \bibinfo {author} {\bibfnamefont {M.~L.}\ \bibnamefont
  {Klein}}, \ and\ \bibinfo {author} {\bibfnamefont {M.}~\bibnamefont
  {Sprik}},\ }\href {\doibase 10.1063/1.2162881} {\bibfield  {journal}
  {\bibinfo  {journal} {J.~Chem. Phys.}\ }\textbf {\bibinfo {volume} {124}},\
  \bibinfo {pages} {064507} (\bibinfo {year} {2006})}\BibitemShut {NoStop}%
\bibitem [{\citenamefont {Krapf}, \citenamefont {Weber},\ and\ \citenamefont
  {Koslowski}(2012)}]{Krapf2012Marcus}%
  \BibitemOpen
  \bibfield  {author} {\bibinfo {author} {\bibfnamefont {S.}~\bibnamefont
  {Krapf}}, \bibinfo {author} {\bibfnamefont {S.}~\bibnamefont {Weber}}, \ and\
  \bibinfo {author} {\bibfnamefont {T.}~\bibnamefont {Koslowski}},\ }\href
  {\doibase 10.1039/c2cp40793k} {\bibfield  {journal} {\bibinfo  {journal}
  {Phys. Chem. Chem. Phys.}\ }\textbf {\bibinfo {volume} {14}},\ \bibinfo
  {pages} {11518} (\bibinfo {year} {2012})}\BibitemShut {NoStop}%
\bibitem [{\citenamefont {Krapf}, \citenamefont {Koslowski},\ and\
  \citenamefont {Steinbrecher}(2010)}]{Krapf2010Marcus}%
  \BibitemOpen
  \bibfield  {author} {\bibinfo {author} {\bibfnamefont {S.}~\bibnamefont
  {Krapf}}, \bibinfo {author} {\bibfnamefont {T.}~\bibnamefont {Koslowski}}, \
  and\ \bibinfo {author} {\bibfnamefont {T.}~\bibnamefont {Steinbrecher}},\
  }\href {\doibase 10.1039/c000876a} {\bibfield  {journal} {\bibinfo  {journal}
  {Phys. Chem. Chem. Phys.}\ }\textbf {\bibinfo {volume} {12}},\ \bibinfo
  {pages} {9516} (\bibinfo {year} {2010})}\BibitemShut {NoStop}%
\bibitem [{\citenamefont {Garg}, \citenamefont {Onuchic},\ and\ \citenamefont
  {Ambegaokar}(1985)}]{Garg1985spinboson}%
  \BibitemOpen
  \bibfield  {author} {\bibinfo {author} {\bibfnamefont {A.}~\bibnamefont
  {Garg}}, \bibinfo {author} {\bibfnamefont {J.~N.}\ \bibnamefont {Onuchic}}, \
  and\ \bibinfo {author} {\bibfnamefont {V.}~\bibnamefont {Ambegaokar}},\
  }\href {\doibase 10.1063/1.449017} {\bibfield  {journal} {\bibinfo  {journal}
  {J.~Chem. Phys.}\ }\textbf {\bibinfo {volume} {83}},\ \bibinfo {pages} {4491}
  (\bibinfo {year} {1985})}\BibitemShut {NoStop}%
\bibitem [{\citenamefont {Ulstrup}(1979)}]{UlstrupBook}%
  \BibitemOpen
  \bibfield  {author} {\bibinfo {author} {\bibfnamefont {J.}~\bibnamefont
  {Ulstrup}},\ }\href@noop {} {\emph {\bibinfo {title} {Charge Transfer
  Processes in Condensed Media}}}\ (\bibinfo  {publisher} {Springer-Verlag},\
  \bibinfo {address} {Berlin},\ \bibinfo {year} {1979})\BibitemShut {NoStop}%
\bibitem [{\citenamefont {Siders}\ and\ \citenamefont
  {Marcus}(1981{\natexlab{a}})}]{Siders1981quantum}%
  \BibitemOpen
  \bibfield  {author} {\bibinfo {author} {\bibfnamefont {P.}~\bibnamefont
  {Siders}}\ and\ \bibinfo {author} {\bibfnamefont {R.}~\bibnamefont
  {Marcus}},\ }\href {\doibase 10.1021/ja00394a003} {\bibfield  {journal}
  {\bibinfo  {journal} {J.~Am. Chem. Soc.}\ }\textbf {\bibinfo {volume}
  {103}},\ \bibinfo {pages} {741} (\bibinfo {year}
  {1981}{\natexlab{a}})}\BibitemShut {NoStop}%
\bibitem [{\citenamefont {Siders}\ and\ \citenamefont
  {Marcus}(1981{\natexlab{b}})}]{Siders1981inverted}%
  \BibitemOpen
  \bibfield  {author} {\bibinfo {author} {\bibfnamefont {P.}~\bibnamefont
  {Siders}}\ and\ \bibinfo {author} {\bibfnamefont {R.}~\bibnamefont
  {Marcus}},\ }\href {\doibase 10.1021/ja00394a004} {\bibfield  {journal}
  {\bibinfo  {journal} {J.~Am. Chem. Soc.}\ }\textbf {\bibinfo {volume}
  {103}},\ \bibinfo {pages} {748} (\bibinfo {year}
  {1981}{\natexlab{b}})}\BibitemShut {NoStop}%
\bibitem [{\citenamefont {Bader}, \citenamefont {Kuharski},\ and\ \citenamefont
  {Chandler}(1990)}]{Bader1990golden}%
  \BibitemOpen
  \bibfield  {author} {\bibinfo {author} {\bibfnamefont {J.~S.}\ \bibnamefont
  {Bader}}, \bibinfo {author} {\bibfnamefont {R.~A.}\ \bibnamefont {Kuharski}},
  \ and\ \bibinfo {author} {\bibfnamefont {D.}~\bibnamefont {Chandler}},\
  }\href {\doibase 10.1063/1.459596} {\bibfield  {journal} {\bibinfo  {journal}
  {J.~Chem. Phys.}\ }\textbf {\bibinfo {volume} {93}},\ \bibinfo {pages} {230}
  (\bibinfo {year} {1990})}\BibitemShut {NoStop}%
\bibitem [{\citenamefont {Topaler}\ and\ \citenamefont
  {Makri}(1996)}]{Topaler1996nonadiabatic}%
  \BibitemOpen
  \bibfield  {author} {\bibinfo {author} {\bibfnamefont {M.}~\bibnamefont
  {Topaler}}\ and\ \bibinfo {author} {\bibfnamefont {N.}~\bibnamefont
  {Makri}},\ }\href {\doibase 10.1021/jp951673k} {\bibfield  {journal}
  {\bibinfo  {journal} {J.~Phys. Chem.}\ }\textbf {\bibinfo {volume} {100}},\
  \bibinfo {pages} {4430} (\bibinfo {year} {1996})}\BibitemShut {NoStop}%
\bibitem [{\citenamefont {Wang}, \citenamefont {Skinner},\ and\ \citenamefont
  {Thoss}(2006)}]{Wang2006flux}%
  \BibitemOpen
  \bibfield  {author} {\bibinfo {author} {\bibfnamefont {H.}~\bibnamefont
  {Wang}}, \bibinfo {author} {\bibfnamefont {D.~E.}\ \bibnamefont {Skinner}}, \
  and\ \bibinfo {author} {\bibfnamefont {M.}~\bibnamefont {Thoss}},\ }\href
  {\doibase 10.1063/1.2363195} {\bibfield  {journal} {\bibinfo  {journal}
  {J.~Chem. Phys.}\ }\textbf {\bibinfo {volume} {125}},\ \bibinfo {pages}
  {174502} (\bibinfo {year} {2006})}\BibitemShut {NoStop}%
\bibitem [{\citenamefont {Tang}(1994{\natexlab{a}})}]{Tang1994asym}%
  \BibitemOpen
  \bibfield  {author} {\bibinfo {author} {\bibfnamefont {J.}~\bibnamefont
  {Tang}},\ }\href {\doibase 10.1016/0301-0104(94)00254-1} {\bibfield
  {journal} {\bibinfo  {journal} {Chem. Phys.}\ }\textbf {\bibinfo {volume}
  {188}},\ \bibinfo {pages} {143} (\bibinfo {year}
  {1994}{\natexlab{a}})}\BibitemShut {NoStop}%
\bibitem [{\citenamefont {Tang}(1994{\natexlab{b}})}]{Tang1994anharmonic}%
  \BibitemOpen
  \bibfield  {author} {\bibinfo {author} {\bibfnamefont {J.}~\bibnamefont
  {Tang}},\ }\href {\doibase 10.1016/0301-0104(93)E0346-W} {\bibfield
  {journal} {\bibinfo  {journal} {Chem. Phys.}\ }\textbf {\bibinfo {volume}
  {179}},\ \bibinfo {pages} {105} (\bibinfo {year}
  {1994}{\natexlab{b}})}\BibitemShut {NoStop}%
\bibitem [{\citenamefont {Miller}(1975)}]{Miller1975semiclassical}%
  \BibitemOpen
  \bibfield  {author} {\bibinfo {author} {\bibfnamefont {W.~H.}\ \bibnamefont
  {Miller}},\ }\href {\doibase 10.1063/1.430676} {\bibfield  {journal}
  {\bibinfo  {journal} {J.~Chem. Phys.}\ }\textbf {\bibinfo {volume} {62}},\
  \bibinfo {pages} {1899} (\bibinfo {year} {1975})}\BibitemShut {NoStop}%
\bibitem [{\citenamefont {Coleman}(1977)}]{Coleman1977ImF}%
  \BibitemOpen
  \bibfield  {author} {\bibinfo {author} {\bibfnamefont {S.}~\bibnamefont
  {Coleman}},\ }\href {\doibase 10.1103/PhysRevD.15.2929} {\bibfield  {journal}
  {\bibinfo  {journal} {Phys. Rev. D}\ }\textbf {\bibinfo {volume} {15}},\
  \bibinfo {pages} {2929} (\bibinfo {year} {1977})}\BibitemShut {NoStop}%
\bibitem [{\citenamefont {Callan}\ and\ \citenamefont
  {Coleman}(1977)}]{Callan1977ImF}%
  \BibitemOpen
  \bibfield  {author} {\bibinfo {author} {\bibfnamefont {C.~G.}\ \bibnamefont
  {Callan}, \bibfnamefont {Jr}}\ and\ \bibinfo {author} {\bibfnamefont
  {S.}~\bibnamefont {Coleman}},\ }\href {\doibase 10.1103/PhysRevD.16.1762}
  {\bibfield  {journal} {\bibinfo  {journal} {Phys. Rev. D}\ }\textbf {\bibinfo
  {volume} {16}},\ \bibinfo {pages} {1762} (\bibinfo {year}
  {1977})}\BibitemShut {NoStop}%
\bibitem [{\citenamefont {Affleck}(1981)}]{Affleck1981ImF}%
  \BibitemOpen
  \bibfield  {author} {\bibinfo {author} {\bibfnamefont {I.}~\bibnamefont
  {Affleck}},\ }\href {\doibase 10.1103/PhysRevLett.46.388} {\bibfield
  {journal} {\bibinfo  {journal} {Phys. Rev. Lett.}\ }\textbf {\bibinfo
  {volume} {46}},\ \bibinfo {pages} {388} (\bibinfo {year} {1981})}\BibitemShut
  {NoStop}%
\bibitem [{\citenamefont {Benderskii}, \citenamefont {Makarov},\ and\
  \citenamefont {Wight}(1994)}]{Benderskii}%
  \BibitemOpen
  \bibfield  {author} {\bibinfo {author} {\bibfnamefont {V.~A.}\ \bibnamefont
  {Benderskii}}, \bibinfo {author} {\bibfnamefont {D.~E.}\ \bibnamefont
  {Makarov}}, \ and\ \bibinfo {author} {\bibfnamefont {C.~A.}\ \bibnamefont
  {Wight}},\ }\href@noop {} {\emph {\bibinfo {title} {Chemical Dynamics at Low
  Temperatures}}},\ \bibinfo {series} {Adv. Chem. Phys.}, Vol.~\bibinfo
  {volume} {88}\ (\bibinfo  {publisher} {Wiley},\ \bibinfo {address} {New
  York},\ \bibinfo {year} {1994})\BibitemShut {NoStop}%
\bibitem [{\citenamefont {Andersson}\ \emph {et~al.}(2009)\citenamefont
  {Andersson}, \citenamefont {Nyman}, \citenamefont {Arnaldsson}, \citenamefont
  {Manthe},\ and\ \citenamefont {J{\'o}nsson}}]{Andersson2009Hmethane}%
  \BibitemOpen
  \bibfield  {author} {\bibinfo {author} {\bibfnamefont {S.}~\bibnamefont
  {Andersson}}, \bibinfo {author} {\bibfnamefont {G.}~\bibnamefont {Nyman}},
  \bibinfo {author} {\bibfnamefont {A.}~\bibnamefont {Arnaldsson}}, \bibinfo
  {author} {\bibfnamefont {U.}~\bibnamefont {Manthe}}, \ and\ \bibinfo {author}
  {\bibfnamefont {H.}~\bibnamefont {J{\'o}nsson}},\ }\href {\doibase
  10.1021/jp811070w} {\bibfield  {journal} {\bibinfo  {journal} {J.~Phys.
  Chem.~A}\ }\textbf {\bibinfo {volume} {113}},\ \bibinfo {pages} {4468}
  (\bibinfo {year} {2009})}\BibitemShut {NoStop}%
\bibitem [{\citenamefont {Richardson}\ and\ \citenamefont
  {Althorpe}(2009)}]{RPinst}%
  \BibitemOpen
  \bibfield  {author} {\bibinfo {author} {\bibfnamefont {J.~O.}\ \bibnamefont
  {Richardson}}\ and\ \bibinfo {author} {\bibfnamefont {S.~C.}\ \bibnamefont
  {Althorpe}},\ }\href {\doibase 10.1063/1.3267318} {\bibfield  {journal}
  {\bibinfo  {journal} {J.~Chem. Phys.}\ }\textbf {\bibinfo {volume} {131}},\
  \bibinfo {pages} {214106} (\bibinfo {year} {2009})}\BibitemShut {NoStop}%
\bibitem [{\citenamefont {Althorpe}(2011)}]{Althorpe2011ImF}%
  \BibitemOpen
  \bibfield  {author} {\bibinfo {author} {\bibfnamefont {S.~C.}\ \bibnamefont
  {Althorpe}},\ }\href {\doibase 10.1063/1.3563045} {\bibfield  {journal}
  {\bibinfo  {journal} {J.~Chem. Phys.}\ }\textbf {\bibinfo {volume} {134}},\
  \bibinfo {pages} {114104} (\bibinfo {year} {2011})}\BibitemShut {NoStop}%
\bibitem [{\citenamefont {Richardson}(2016{\natexlab{a}})}]{AdiabaticGreens}%
  \BibitemOpen
  \bibfield  {author} {\bibinfo {author} {\bibfnamefont {J.~O.}\ \bibnamefont
  {Richardson}},\ }\href {\doibase 10.1063/1.4943866} {\bibfield  {journal}
  {\bibinfo  {journal} {J.~Chem. Phys.}\ }\textbf {\bibinfo {volume} {144}},\
  \bibinfo {pages} {114106} (\bibinfo {year} {2016}{\natexlab{a}})},\ \Eprint
  {http://arxiv.org/abs/1512.04292} {arXiv:1512.04292 [physics.chem-ph]}
  \BibitemShut {NoStop}%
\bibitem [{\citenamefont {Richardson}(2016{\natexlab{b}})}]{faraday}%
  \BibitemOpen
  \bibfield  {author} {\bibinfo {author} {\bibfnamefont {J.~O.}\ \bibnamefont
  {Richardson}},\ }\href {\doibase 10.1039/C6FD00119J} {\bibfield  {journal}
  {\bibinfo  {journal} {Faraday Discuss.}\ }\textbf {\bibinfo {volume} {195}},\
  \bibinfo {pages} {49} (\bibinfo {year} {2016}{\natexlab{b}})}\BibitemShut
  {NoStop}%
\bibitem [{\citenamefont {K\"astner}(2014)}]{Kaestner2014review}%
  \BibitemOpen
  \bibfield  {author} {\bibinfo {author} {\bibfnamefont {J.}~\bibnamefont
  {K\"astner}},\ }\href {\doibase 10.1002/wcms.1165} {\bibfield  {journal}
  {\bibinfo  {journal} {WIREs Comput. Mol. Sci.}\ }\textbf {\bibinfo {volume}
  {4}},\ \bibinfo {pages} {158} (\bibinfo {year} {2014})}\BibitemShut {NoStop}%
\bibitem [{\citenamefont {Ceotto}(2012)}]{Ceotto2012instanton}%
  \BibitemOpen
  \bibfield  {author} {\bibinfo {author} {\bibfnamefont {M.}~\bibnamefont
  {Ceotto}},\ }\href {\doibase 10.1080/00268976.2012.663943} {\bibfield
  {journal} {\bibinfo  {journal} {Mol. Phys.}\ }\textbf {\bibinfo {volume}
  {110}},\ \bibinfo {pages} {547} (\bibinfo {year} {2012})}\BibitemShut
  {NoStop}%
\bibitem [{\citenamefont {Richardson}, \citenamefont {Bauer},\ and\
  \citenamefont {Thoss}(2015)}]{GoldenGreens}%
  \BibitemOpen
  \bibfield  {author} {\bibinfo {author} {\bibfnamefont {J.~O.}\ \bibnamefont
  {Richardson}}, \bibinfo {author} {\bibfnamefont {R.}~\bibnamefont {Bauer}}, \
  and\ \bibinfo {author} {\bibfnamefont {M.}~\bibnamefont {Thoss}},\ }\href
  {\doibase 10.1063/1.4932361} {\bibfield  {journal} {\bibinfo  {journal}
  {J.~Chem. Phys.}\ }\textbf {\bibinfo {volume} {143}},\ \bibinfo {pages}
  {134115} (\bibinfo {year} {2015})},\ \Eprint
  {http://arxiv.org/abs/1508.04919} {arXiv:1508.04919 [physics.chem-ph]}
  \BibitemShut {NoStop}%
\bibitem [{\citenamefont {Richardson}(2015)}]{GoldenRPI}%
  \BibitemOpen
  \bibfield  {author} {\bibinfo {author} {\bibfnamefont {J.~O.}\ \bibnamefont
  {Richardson}},\ }\href {\doibase 10.1063/1.4932362} {\bibfield  {journal}
  {\bibinfo  {journal} {J.~Chem. Phys.}\ }\textbf {\bibinfo {volume} {143}},\
  \bibinfo {pages} {134116} (\bibinfo {year} {2015})},\ \Eprint
  {http://arxiv.org/abs/1508.05195} {arXiv:1508.05195 [physics.chem-ph]}
  \BibitemShut {NoStop}%
\bibitem [{\citenamefont
  {Wolynes}(1987{\natexlab{a}})}]{Wolynes1987nonadiabatic}%
  \BibitemOpen
  \bibfield  {author} {\bibinfo {author} {\bibfnamefont {P.~G.}\ \bibnamefont
  {Wolynes}},\ }\href {\doibase 10.1063/1.453440} {\bibfield  {journal}
  {\bibinfo  {journal} {J.~Chem. Phys.}\ }\textbf {\bibinfo {volume} {87}},\
  \bibinfo {pages} {6559} (\bibinfo {year} {1987}{\natexlab{a}})}\BibitemShut
  {NoStop}%
\bibitem [{\citenamefont {Zheng}, \citenamefont {McCammon},\ and\ \citenamefont
  {Wolynes}(1989)}]{Zheng1989ET}%
  \BibitemOpen
  \bibfield  {author} {\bibinfo {author} {\bibfnamefont {C.}~\bibnamefont
  {Zheng}}, \bibinfo {author} {\bibfnamefont {J.~A.}\ \bibnamefont {McCammon}},
  \ and\ \bibinfo {author} {\bibfnamefont {P.~G.}\ \bibnamefont {Wolynes}},\
  }\href {\doibase 10.1073/pnas.86.17.6441} {\bibfield  {journal} {\bibinfo
  {journal} {P. Natl. Acad. Sci. USA}\ }\textbf {\bibinfo {volume} {86}},\
  \bibinfo {pages} {6441} (\bibinfo {year} {1989})}\BibitemShut {NoStop}%
\bibitem [{\citenamefont {Zheng}, \citenamefont {McCammon},\ and\ \citenamefont
  {Wolynes}(1991)}]{Zheng1991ET}%
  \BibitemOpen
  \bibfield  {author} {\bibinfo {author} {\bibfnamefont {C.}~\bibnamefont
  {Zheng}}, \bibinfo {author} {\bibfnamefont {J.~A.}\ \bibnamefont {McCammon}},
  \ and\ \bibinfo {author} {\bibfnamefont {P.~G.}\ \bibnamefont {Wolynes}},\
  }\href {\doibase 10.1016/0301-0104(91)87070-C} {\bibfield  {journal}
  {\bibinfo  {journal} {Chem. Phys.}\ }\textbf {\bibinfo {volume} {158}},\
  \bibinfo {pages} {261} (\bibinfo {year} {1991})}\BibitemShut {NoStop}%
\bibitem [{\citenamefont {Richardson}\ and\ \citenamefont
  {Thoss}(2014)}]{nonoscillatory}%
  \BibitemOpen
  \bibfield  {author} {\bibinfo {author} {\bibfnamefont {J.~O.}\ \bibnamefont
  {Richardson}}\ and\ \bibinfo {author} {\bibfnamefont {M.}~\bibnamefont
  {Thoss}},\ }\href {\doibase 10.1063/1.4892865} {\bibfield  {journal}
  {\bibinfo  {journal} {J.~Chem. Phys.}\ }\textbf {\bibinfo {volume} {141}},\
  \bibinfo {pages} {074106} (\bibinfo {year} {2014})},\ \Eprint
  {http://arxiv.org/abs/1406.3144} {arXiv:1406.3144 [physics.chem-ph]}
  \BibitemShut {NoStop}%
\bibitem [{\citenamefont {Rips}\ and\ \citenamefont
  {Pollak}(1995)}]{Rips1995ET}%
  \BibitemOpen
  \bibfield  {author} {\bibinfo {author} {\bibfnamefont {I.}~\bibnamefont
  {Rips}}\ and\ \bibinfo {author} {\bibfnamefont {E.}~\bibnamefont {Pollak}},\
  }\href {\doibase 10.1063/1.470209} {\bibfield  {journal} {\bibinfo  {journal}
  {J.~Chem. Phys.}\ }\textbf {\bibinfo {volume} {103}},\ \bibinfo {pages}
  {7912} (\bibinfo {year} {1995})}\BibitemShut {NoStop}%
\bibitem [{\citenamefont {M{\"u}hlbacher}\ and\ \citenamefont
  {Egger}(2003)}]{Muehlbacher2003spinboson}%
  \BibitemOpen
  \bibfield  {author} {\bibinfo {author} {\bibfnamefont {L.}~\bibnamefont
  {M{\"u}hlbacher}}\ and\ \bibinfo {author} {\bibfnamefont {R.}~\bibnamefont
  {Egger}},\ }\href {\doibase 10.1063/1.1523014} {\bibfield  {journal}
  {\bibinfo  {journal} {J.~Chem. Phys.}\ }\textbf {\bibinfo {volume} {118}},\
  \bibinfo {pages} {179} (\bibinfo {year} {2003})}\BibitemShut {NoStop}%
\bibitem [{\citenamefont {Casado-Pascual}\ \emph {et~al.}(2003)\citenamefont
  {Casado-Pascual}, \citenamefont {Morillo}, \citenamefont {Goychuk},\ and\
  \citenamefont {H{\"a}nggi}}]{CasadoPascual2003ET}%
  \BibitemOpen
  \bibfield  {author} {\bibinfo {author} {\bibfnamefont {J.}~\bibnamefont
  {Casado-Pascual}}, \bibinfo {author} {\bibfnamefont {M.}~\bibnamefont
  {Morillo}}, \bibinfo {author} {\bibfnamefont {I.}~\bibnamefont {Goychuk}}, \
  and\ \bibinfo {author} {\bibfnamefont {P.}~\bibnamefont {H{\"a}nggi}},\
  }\href {\doibase 10.1063/1.1525799} {\bibfield  {journal} {\bibinfo
  {journal} {J.~Chem. Phys.}\ }\textbf {\bibinfo {volume} {118}},\ \bibinfo
  {pages} {291} (\bibinfo {year} {2003})}\BibitemShut {NoStop}%
\bibitem [{\citenamefont {Barzykin}\ \emph {et~al.}(2002)\citenamefont
  {Barzykin}, \citenamefont {Frantsuzov}, \citenamefont {Seki},\ and\
  \citenamefont {Tachiya}}]{Barzykin2002ET}%
  \BibitemOpen
  \bibfield  {author} {\bibinfo {author} {\bibfnamefont {A.}~\bibnamefont
  {Barzykin}}, \bibinfo {author} {\bibfnamefont {P.}~\bibnamefont
  {Frantsuzov}}, \bibinfo {author} {\bibfnamefont {K.}~\bibnamefont {Seki}}, \
  and\ \bibinfo {author} {\bibfnamefont {M.}~\bibnamefont {Tachiya}},\
  }\href@noop {} {\bibfield  {journal} {\bibinfo  {journal} {Adv. Chem. Phys.}\
  }\textbf {\bibinfo {volume} {123}},\ \bibinfo {pages} {511} (\bibinfo {year}
  {2002})}\BibitemShut {NoStop}%
\bibitem [{\citenamefont {Spencer}\ \emph {et~al.}(2016)\citenamefont
  {Spencer}, \citenamefont {Scalfi}, \citenamefont {Carof},\ and\ \citenamefont
  {Blumberger}}]{Spencer2016Faraday}%
  \BibitemOpen
  \bibfield  {author} {\bibinfo {author} {\bibfnamefont {J.}~\bibnamefont
  {Spencer}}, \bibinfo {author} {\bibfnamefont {L.}~\bibnamefont {Scalfi}},
  \bibinfo {author} {\bibfnamefont {A.}~\bibnamefont {Carof}}, \ and\ \bibinfo
  {author} {\bibfnamefont {J.}~\bibnamefont {Blumberger}},\ }\href {\doibase
  10.1039/c6fd00107f} {\bibfield  {journal} {\bibinfo  {journal} {Faraday
  Discuss.}\ }\textbf {\bibinfo {volume} {195}},\ \bibinfo {pages} {215}
  (\bibinfo {year} {2016})}\BibitemShut {NoStop}%
\bibitem [{\citenamefont {Cao}, \citenamefont {Minichino},\ and\ \citenamefont
  {Voth}(1995)}]{Cao1995nonadiabatic}%
  \BibitemOpen
  \bibfield  {author} {\bibinfo {author} {\bibfnamefont {J.}~\bibnamefont
  {Cao}}, \bibinfo {author} {\bibfnamefont {C.}~\bibnamefont {Minichino}}, \
  and\ \bibinfo {author} {\bibfnamefont {G.~A.}\ \bibnamefont {Voth}},\ }\href
  {\doibase 10.1063/1.469762} {\bibfield  {journal} {\bibinfo  {journal}
  {J.~Chem. Phys.}\ }\textbf {\bibinfo {volume} {103}},\ \bibinfo {pages}
  {1391} (\bibinfo {year} {1995})}\BibitemShut {NoStop}%
\bibitem [{\citenamefont {Cao}\ and\ \citenamefont
  {Voth}(1997)}]{Cao1997nonadiabatic}%
  \BibitemOpen
  \bibfield  {author} {\bibinfo {author} {\bibfnamefont {J.}~\bibnamefont
  {Cao}}\ and\ \bibinfo {author} {\bibfnamefont {G.~A.}\ \bibnamefont {Voth}},\
  }\href {\doibase 10.1063/1.474123} {\bibfield  {journal} {\bibinfo  {journal}
  {J.~Chem. Phys.}\ }\textbf {\bibinfo {volume} {106}},\ \bibinfo {pages}
  {1769} (\bibinfo {year} {1997})}\BibitemShut {NoStop}%
\bibitem [{\citenamefont {Schwieters}\ and\ \citenamefont
  {Voth}(1998)}]{Schwieters1998diabatic}%
  \BibitemOpen
  \bibfield  {author} {\bibinfo {author} {\bibfnamefont {C.~D.}\ \bibnamefont
  {Schwieters}}\ and\ \bibinfo {author} {\bibfnamefont {G.~A.}\ \bibnamefont
  {Voth}},\ }\href {\doibase 10.1063/1.475467} {\bibfield  {journal} {\bibinfo
  {journal} {J.~Chem. Phys.}\ }\textbf {\bibinfo {volume} {108}},\ \bibinfo
  {pages} {1055} (\bibinfo {year} {1998})}\BibitemShut {NoStop}%
\bibitem [{\citenamefont {Zhu}\ and\ \citenamefont
  {Nakamura}(1994)}]{Zhu1994ZN}%
  \BibitemOpen
  \bibfield  {author} {\bibinfo {author} {\bibfnamefont {C.}~\bibnamefont
  {Zhu}}\ and\ \bibinfo {author} {\bibfnamefont {H.}~\bibnamefont {Nakamura}},\
  }\href {\doibase 10.1063/1.467877} {\bibfield  {journal} {\bibinfo  {journal}
  {J.~Chem. Phys.}\ }\textbf {\bibinfo {volume} {101}},\ \bibinfo {pages}
  {10630} (\bibinfo {year} {1994})}\BibitemShut {NoStop}%
\bibitem [{\citenamefont {Zhu}\ and\ \citenamefont
  {Nakamura}(1995)}]{Zhu1995ZN}%
  \BibitemOpen
  \bibfield  {author} {\bibinfo {author} {\bibfnamefont {C.}~\bibnamefont
  {Zhu}}\ and\ \bibinfo {author} {\bibfnamefont {H.}~\bibnamefont {Nakamura}},\
  }\href {\doibase 10.1063/1.469057} {\bibfield  {journal} {\bibinfo  {journal}
  {J.~Chem. Phys.}\ }\textbf {\bibinfo {volume} {102}},\ \bibinfo {pages}
  {7448} (\bibinfo {year} {1995})}\BibitemShut {NoStop}%
\bibitem [{\citenamefont {Zhao}, \citenamefont {Mil'nikov},\ and\ \citenamefont
  {Nakamura}(2004)}]{Zhao2004nonadiabatic}%
  \BibitemOpen
  \bibfield  {author} {\bibinfo {author} {\bibfnamefont {Y.}~\bibnamefont
  {Zhao}}, \bibinfo {author} {\bibfnamefont {G.}~\bibnamefont {Mil'nikov}}, \
  and\ \bibinfo {author} {\bibfnamefont {H.}~\bibnamefont {Nakamura}},\ }\href
  {\doibase 10.1063/1.1801971} {\bibfield  {journal} {\bibinfo  {journal}
  {J.~Chem. Phys.}\ }\textbf {\bibinfo {volume} {121}},\ \bibinfo {pages}
  {8854} (\bibinfo {year} {2004})}\BibitemShut {NoStop}%
\bibitem [{\citenamefont {Miller}(1974)}]{Miller1974QTST}%
  \BibitemOpen
  \bibfield  {author} {\bibinfo {author} {\bibfnamefont {W.~H.}\ \bibnamefont
  {Miller}},\ }\href {\doibase 10.1063/1.1682181} {\bibfield  {journal}
  {\bibinfo  {journal} {J.~Chem. Phys.}\ }\textbf {\bibinfo {volume} {61}},\
  \bibinfo {pages} {1823} (\bibinfo {year} {1974})}\BibitemShut {NoStop}%
\bibitem [{\citenamefont {Miller}, \citenamefont {Schwartz},\ and\
  \citenamefont {Tromp}(1983)}]{Miller1983rate}%
  \BibitemOpen
  \bibfield  {author} {\bibinfo {author} {\bibfnamefont {W.~H.}\ \bibnamefont
  {Miller}}, \bibinfo {author} {\bibfnamefont {S.~D.}\ \bibnamefont
  {Schwartz}}, \ and\ \bibinfo {author} {\bibfnamefont {J.~W.}\ \bibnamefont
  {Tromp}},\ }\href {\doibase 10.1063/1.445581} {\bibfield  {journal} {\bibinfo
   {journal} {J.~Chem. Phys.}\ }\textbf {\bibinfo {volume} {79}},\ \bibinfo
  {pages} {4889} (\bibinfo {year} {1983})}\BibitemShut {NoStop}%
\bibitem [{\citenamefont {Chandler}(1987)}]{ChandlerGreen}%
  \BibitemOpen
  \bibfield  {author} {\bibinfo {author} {\bibfnamefont {D.}~\bibnamefont
  {Chandler}},\ }\href@noop {} {\emph {\bibinfo {title} {Introduction to Modern
  Statistical Mechanics}}}\ (\bibinfo  {publisher} {Oxford University Press},\
  \bibinfo {address} {New York},\ \bibinfo {year} {1987})\BibitemShut {NoStop}%
\bibitem [{\citenamefont {Miller}(1971)}]{Miller1971density}%
  \BibitemOpen
  \bibfield  {author} {\bibinfo {author} {\bibfnamefont {W.~H.}\ \bibnamefont
  {Miller}},\ }\href {\doibase 10.1063/1.1676560} {\bibfield  {journal}
  {\bibinfo  {journal} {J.~Chem. Phys.}\ }\textbf {\bibinfo {volume} {55}},\
  \bibinfo {pages} {3146} (\bibinfo {year} {1971})}\BibitemShut {NoStop}%
\bibitem [{\citenamefont {Gutzwiller}(1990)}]{GutzwillerBook}%
  \BibitemOpen
  \bibfield  {author} {\bibinfo {author} {\bibfnamefont {M.~C.}\ \bibnamefont
  {Gutzwiller}},\ }\href@noop {} {\emph {\bibinfo {title} {Chaos in Classical
  and Quantum Mechanics}}}\ (\bibinfo  {publisher} {Springer-Verlag},\ \bibinfo
  {address} {New York},\ \bibinfo {year} {1990})\BibitemShut {NoStop}%
\bibitem [{\citenamefont {Kleinert}(2009)}]{Kleinert}%
  \BibitemOpen
  \bibfield  {author} {\bibinfo {author} {\bibfnamefont {H.}~\bibnamefont
  {Kleinert}},\ }\href@noop {} {\emph {\bibinfo {title} {Path Integrals in
  Quantum Mechanics, Statistics, Polymer Physics and Financial Markets}}},\
  \bibinfo {edition} {5th}\ ed.\ (\bibinfo  {publisher} {World Scientific},\
  \bibinfo {address} {Singapore},\ \bibinfo {year} {2009})\BibitemShut
  {NoStop}%
\bibitem [{\citenamefont {Miller}\ \emph {et~al.}(2003)\citenamefont {Miller},
  \citenamefont {Zhao}, \citenamefont {Ceotto},\ and\ \citenamefont
  {Yang}}]{Miller2003QI}%
  \BibitemOpen
  \bibfield  {author} {\bibinfo {author} {\bibfnamefont {W.~H.}\ \bibnamefont
  {Miller}}, \bibinfo {author} {\bibfnamefont {Y.}~\bibnamefont {Zhao}},
  \bibinfo {author} {\bibfnamefont {M.}~\bibnamefont {Ceotto}}, \ and\ \bibinfo
  {author} {\bibfnamefont {S.}~\bibnamefont {Yang}},\ }\href {\doibase
  10.1063/1.1580110} {\bibfield  {journal} {\bibinfo  {journal} {J.~Chem.
  Phys.}\ }\textbf {\bibinfo {volume} {119}},\ \bibinfo {pages} {1329}
  (\bibinfo {year} {2003})}\BibitemShut {NoStop}%
\bibitem [{\citenamefont {Van{\'\i}{\v{c}}ek}\ \emph
  {et~al.}(2005)\citenamefont {Van{\'\i}{\v{c}}ek}, \citenamefont {Miller},
  \citenamefont {Castillo},\ and\ \citenamefont {Aoiz}}]{Vanicek2005QI}%
  \BibitemOpen
  \bibfield  {author} {\bibinfo {author} {\bibfnamefont {J.}~\bibnamefont
  {Van{\'\i}{\v{c}}ek}}, \bibinfo {author} {\bibfnamefont {W.~H.}\ \bibnamefont
  {Miller}}, \bibinfo {author} {\bibfnamefont {J.~F.}\ \bibnamefont
  {Castillo}}, \ and\ \bibinfo {author} {\bibfnamefont {F.~J.}\ \bibnamefont
  {Aoiz}},\ }\href {\doibase 10.1063/1.1946740} {\bibfield  {journal} {\bibinfo
   {journal} {J.~Chem. Phys.}\ }\textbf {\bibinfo {volume} {123}},\ \bibinfo
  {pages} {054108} (\bibinfo {year} {2005})}\BibitemShut {NoStop}%
\bibitem [{\citenamefont {Beyer}\ \emph {et~al.}(2016)\citenamefont {Beyer},
  \citenamefont {Richardson}, \citenamefont {Knowles}, \citenamefont {Rommel},\
  and\ \citenamefont {Althorpe}}]{HCH4}%
  \BibitemOpen
  \bibfield  {author} {\bibinfo {author} {\bibfnamefont {A.~N.}\ \bibnamefont
  {Beyer}}, \bibinfo {author} {\bibfnamefont {J.~O.}\ \bibnamefont
  {Richardson}}, \bibinfo {author} {\bibfnamefont {P.~J.}\ \bibnamefont
  {Knowles}}, \bibinfo {author} {\bibfnamefont {J.}~\bibnamefont {Rommel}}, \
  and\ \bibinfo {author} {\bibfnamefont {S.~C.}\ \bibnamefont {Althorpe}},\
  }\href {\doibase 10.1021/acs.jpclett.6b02115} {\bibfield  {journal} {\bibinfo
   {journal} {J.~Phys. Chem. Lett.}\ }\textbf {\bibinfo {volume} {7}},\
  \bibinfo {pages} {4374} (\bibinfo {year} {2016})}\BibitemShut {NoStop}%
\bibitem [{\citenamefont {Richardson}\ and\ \citenamefont
  {Althorpe}(2011)}]{tunnel}%
  \BibitemOpen
  \bibfield  {author} {\bibinfo {author} {\bibfnamefont {J.~O.}\ \bibnamefont
  {Richardson}}\ and\ \bibinfo {author} {\bibfnamefont {S.~C.}\ \bibnamefont
  {Althorpe}},\ }\href {\doibase 10.1063/1.3530589} {\bibfield  {journal}
  {\bibinfo  {journal} {J.~Chem. Phys.}\ }\textbf {\bibinfo {volume} {134}},\
  \bibinfo {pages} {054109} (\bibinfo {year} {2011})}\BibitemShut {NoStop}%
\bibitem [{\citenamefont {Richardson}, \citenamefont {Althorpe},\ and\
  \citenamefont {Wales}(2011)}]{water}%
  \BibitemOpen
  \bibfield  {author} {\bibinfo {author} {\bibfnamefont {J.~O.}\ \bibnamefont
  {Richardson}}, \bibinfo {author} {\bibfnamefont {S.~C.}\ \bibnamefont
  {Althorpe}}, \ and\ \bibinfo {author} {\bibfnamefont {D.~J.}\ \bibnamefont
  {Wales}},\ }\href {\doibase 10.1063/1.3640429} {\bibfield  {journal}
  {\bibinfo  {journal} {J.~Chem. Phys.}\ }\textbf {\bibinfo {volume} {135}},\
  \bibinfo {pages} {124109} (\bibinfo {year} {2011})}\BibitemShut {NoStop}%
\bibitem [{\citenamefont {Richardson}\ \emph {et~al.}(2013)\citenamefont
  {Richardson}, \citenamefont {Wales}, \citenamefont {Althorpe}, \citenamefont
  {McLaughlin}, \citenamefont {Viant}, \citenamefont {Shih},\ and\
  \citenamefont {Saykally}}]{octamer}%
  \BibitemOpen
  \bibfield  {author} {\bibinfo {author} {\bibfnamefont {J.~O.}\ \bibnamefont
  {Richardson}}, \bibinfo {author} {\bibfnamefont {D.~J.}\ \bibnamefont
  {Wales}}, \bibinfo {author} {\bibfnamefont {S.~C.}\ \bibnamefont {Althorpe}},
  \bibinfo {author} {\bibfnamefont {R.~P.}\ \bibnamefont {McLaughlin}},
  \bibinfo {author} {\bibfnamefont {M.~R.}\ \bibnamefont {Viant}}, \bibinfo
  {author} {\bibfnamefont {O.}~\bibnamefont {Shih}}, \ and\ \bibinfo {author}
  {\bibfnamefont {R.~J.}\ \bibnamefont {Saykally}},\ }\href {\doibase
  10.1021/jp311306a} {\bibfield  {journal} {\bibinfo  {journal} {J.~Phys.
  Chem.~A}\ }\textbf {\bibinfo {volume} {117}},\ \bibinfo {pages} {6960 }
  (\bibinfo {year} {2013})}\BibitemShut {NoStop}%
\bibitem [{\citenamefont {Richardson}\ \emph {et~al.}(2016)\citenamefont
  {Richardson}, \citenamefont {P{\'e}rez}, \citenamefont {Lobsiger},
  \citenamefont {Reid}, \citenamefont {Temelso}, \citenamefont {Shields},
  \citenamefont {Kisiel}, \citenamefont {Wales}, \citenamefont {Pate},\ and\
  \citenamefont {Althorpe}}]{hexamerprism}%
  \BibitemOpen
  \bibfield  {author} {\bibinfo {author} {\bibfnamefont {J.~O.}\ \bibnamefont
  {Richardson}}, \bibinfo {author} {\bibfnamefont {C.}~\bibnamefont
  {P{\'e}rez}}, \bibinfo {author} {\bibfnamefont {S.}~\bibnamefont {Lobsiger}},
  \bibinfo {author} {\bibfnamefont {A.~A.}\ \bibnamefont {Reid}}, \bibinfo
  {author} {\bibfnamefont {B.}~\bibnamefont {Temelso}}, \bibinfo {author}
  {\bibfnamefont {G.~C.}\ \bibnamefont {Shields}}, \bibinfo {author}
  {\bibfnamefont {Z.}~\bibnamefont {Kisiel}}, \bibinfo {author} {\bibfnamefont
  {D.~J.}\ \bibnamefont {Wales}}, \bibinfo {author} {\bibfnamefont {B.~H.}\
  \bibnamefont {Pate}}, \ and\ \bibinfo {author} {\bibfnamefont {S.~C.}\
  \bibnamefont {Althorpe}},\ }\href {\doibase 10.1126/science.aae0012}
  {\bibfield  {journal} {\bibinfo  {journal} {Science}\ }\textbf {\bibinfo
  {volume} {351}},\ \bibinfo {pages} {1310} (\bibinfo {year}
  {2016})}\BibitemShut {NoStop}%
\bibitem [{\citenamefont {Nichols}\ \emph {et~al.}(1990)\citenamefont
  {Nichols}, \citenamefont {Taylor}, \citenamefont {Schmidt},\ and\
  \citenamefont {Simons}}]{Nichols1990mep}%
  \BibitemOpen
  \bibfield  {author} {\bibinfo {author} {\bibfnamefont {J.}~\bibnamefont
  {Nichols}}, \bibinfo {author} {\bibfnamefont {H.}~\bibnamefont {Taylor}},
  \bibinfo {author} {\bibfnamefont {P.}~\bibnamefont {Schmidt}}, \ and\
  \bibinfo {author} {\bibfnamefont {J.}~\bibnamefont {Simons}},\ }\href
  {\doibase 10.1063/1.458435} {\bibfield  {journal} {\bibinfo  {journal}
  {J.~Chem. Phys.}\ }\textbf {\bibinfo {volume} {92}},\ \bibinfo {pages} {340}
  (\bibinfo {year} {1990})}\BibitemShut {NoStop}%
\bibitem [{\citenamefont {H{\"a}nggi}\ and\ \citenamefont
  {Hontscha}(1991)}]{Haenggi1991instanton}%
  \BibitemOpen
  \bibfield  {author} {\bibinfo {author} {\bibfnamefont {P.}~\bibnamefont
  {H{\"a}nggi}}\ and\ \bibinfo {author} {\bibfnamefont {W.}~\bibnamefont
  {Hontscha}},\ }\href@noop {} {\bibfield  {journal} {\bibinfo  {journal} {Ber.
  Bunsenges. Phys. Chem.}\ }\textbf {\bibinfo {volume} {95}},\ \bibinfo {pages}
  {379} (\bibinfo {year} {1991})}\BibitemShut {NoStop}%
\bibitem [{\citenamefont {Kryvohuz}(2011)}]{Kryvohuz2011rate}%
  \BibitemOpen
  \bibfield  {author} {\bibinfo {author} {\bibfnamefont {M.}~\bibnamefont
  {Kryvohuz}},\ }\href {\doibase 10.1063/1.3565425} {\bibfield  {journal}
  {\bibinfo  {journal} {J.~Chem. Phys.}\ }\textbf {\bibinfo {volume} {134}},\
  \bibinfo {pages} {114103} (\bibinfo {year} {2011})}\BibitemShut {NoStop}%
\bibitem [{\citenamefont {Zener}(1932)}]{Zener1932LZ}%
  \BibitemOpen
  \bibfield  {author} {\bibinfo {author} {\bibfnamefont {C.}~\bibnamefont
  {Zener}},\ }\href {\doibase 10.1098/rspa.1932.0165} {\bibfield  {journal}
  {\bibinfo  {journal} {Proc. R. Soc. Lond. A}\ }\textbf {\bibinfo {volume}
  {137}},\ \bibinfo {pages} {696} (\bibinfo {year} {1932})}\BibitemShut
  {NoStop}%
\bibitem [{\citenamefont {Nitzan}(2006)}]{Nitzan}%
  \BibitemOpen
  \bibfield  {author} {\bibinfo {author} {\bibfnamefont {A.}~\bibnamefont
  {Nitzan}},\ }\href@noop {} {\emph {\bibinfo {title} {Chemical Dynamics in
  Condensed Phases: Relaxation, Transfer, and Reactions in Condensed Molecular
  Systems}}}\ (\bibinfo  {publisher} {Oxford University Press},\ \bibinfo
  {year} {2006})\BibitemShut {NoStop}%
\bibitem [{\citenamefont {Caldeira}\ and\ \citenamefont
  {Leggett}(1983)}]{Caldeira1983dissipation}%
  \BibitemOpen
  \bibfield  {author} {\bibinfo {author} {\bibfnamefont {A.~O.}\ \bibnamefont
  {Caldeira}}\ and\ \bibinfo {author} {\bibfnamefont {A.~J.}\ \bibnamefont
  {Leggett}},\ }\href {\doibase 10.1016/0003-4916(83)90202-6} {\bibfield
  {journal} {\bibinfo  {journal} {Ann. Phys.--New York}\ }\textbf {\bibinfo
  {volume} {149}},\ \bibinfo {pages} {374} (\bibinfo {year}
  {1983})}\BibitemShut {NoStop}%
\bibitem [{\citenamefont {Weiss}(2012)}]{Weiss}%
  \BibitemOpen
  \bibfield  {author} {\bibinfo {author} {\bibfnamefont {U.}~\bibnamefont
  {Weiss}},\ }\href@noop {} {\emph {\bibinfo {title} {Quantum Dissipative
  Systems}}},\ \bibinfo {edition} {4th}\ ed.\ (\bibinfo  {publisher} {World
  Scientific},\ \bibinfo {address} {Singapore},\ \bibinfo {year}
  {2012})\BibitemShut {NoStop}%
\bibitem [{\citenamefont {Craig}\ and\ \citenamefont
  {Manolopoulos}(2005)}]{RPMDrate}%
  \BibitemOpen
  \bibfield  {author} {\bibinfo {author} {\bibfnamefont {I.~R.}\ \bibnamefont
  {Craig}}\ and\ \bibinfo {author} {\bibfnamefont {D.~E.}\ \bibnamefont
  {Manolopoulos}},\ }\href {\doibase 10.1063/1.1850093} {\bibfield  {journal}
  {\bibinfo  {journal} {J.~Chem. Phys.}\ }\textbf {\bibinfo {volume} {122}},\
  \bibinfo {eid} {084106} (\bibinfo {year} {2005}),\
  10.1063/1.1850093}\BibitemShut {NoStop}%
\bibitem [{\citenamefont {Thoss}, \citenamefont {Wang},\ and\ \citenamefont
  {Miller}(2001)}]{Thoss2001hybrid}%
  \BibitemOpen
  \bibfield  {author} {\bibinfo {author} {\bibfnamefont {M.}~\bibnamefont
  {Thoss}}, \bibinfo {author} {\bibfnamefont {H.}~\bibnamefont {Wang}}, \ and\
  \bibinfo {author} {\bibfnamefont {W.~H.}\ \bibnamefont {Miller}},\ }\href
  {\doibase 10.1063/1.1385562} {\bibfield  {journal} {\bibinfo  {journal}
  {J.~Chem. Phys.}\ }\textbf {\bibinfo {volume} {115}},\ \bibinfo {pages}
  {2991} (\bibinfo {year} {2001})}\BibitemShut {NoStop}%
\bibitem [{\citenamefont {Schwerdtfeger}, \citenamefont {Soudackov},\ and\
  \citenamefont {Hammes-Schiffer}(2014)}]{Schwerdtfeger2014ET}%
  \BibitemOpen
  \bibfield  {author} {\bibinfo {author} {\bibfnamefont {C.~A.}\ \bibnamefont
  {Schwerdtfeger}}, \bibinfo {author} {\bibfnamefont {A.~V.}\ \bibnamefont
  {Soudackov}}, \ and\ \bibinfo {author} {\bibfnamefont {S.}~\bibnamefont
  {Hammes-Schiffer}},\ }\href {\doibase 10.1063/1.4855295} {\bibfield
  {journal} {\bibinfo  {journal} {J.~Chem. Phys.}\ }\textbf {\bibinfo {volume}
  {140}},\ \bibinfo {pages} {034113} (\bibinfo {year} {2014})}\BibitemShut
  {NoStop}%
\bibitem [{\citenamefont {Blumberger}\ and\ \citenamefont
  {Sprik}(2006)}]{Blumberger2006RuRu}%
  \BibitemOpen
  \bibfield  {author} {\bibinfo {author} {\bibfnamefont {J.}~\bibnamefont
  {Blumberger}}\ and\ \bibinfo {author} {\bibfnamefont {M.}~\bibnamefont
  {Sprik}},\ }\href {\doibase 10.1007/s00214-005-0058-0} {\bibfield  {journal}
  {\bibinfo  {journal} {Theor. Chem. Acc.}\ }\textbf {\bibinfo {volume}
  {115}},\ \bibinfo {pages} {113} (\bibinfo {year} {2006})}\BibitemShut
  {NoStop}%
\bibitem [{\citenamefont {Wigner}(1938)}]{Wigner1938TST}%
  \BibitemOpen
  \bibfield  {author} {\bibinfo {author} {\bibfnamefont {E.}~\bibnamefont
  {Wigner}},\ }\href@noop {} {\bibfield  {journal} {\bibinfo  {journal} {Trans.
  Faraday Soc.}\ }\textbf {\bibinfo {volume} {34}},\ \bibinfo {pages} {29}
  (\bibinfo {year} {1938})}\BibitemShut {NoStop}%
\bibitem [{\citenamefont {Pollak}(1986)}]{Pollak1986Kramers}%
  \BibitemOpen
  \bibfield  {author} {\bibinfo {author} {\bibfnamefont {E.}~\bibnamefont
  {Pollak}},\ }\href {\doibase 10.1063/1.451294} {\bibfield  {journal}
  {\bibinfo  {journal} {J.~Chem. Phys.}\ }\textbf {\bibinfo {volume} {85}},\
  \bibinfo {pages} {865} (\bibinfo {year} {1986})}\BibitemShut {NoStop}%
\bibitem [{\citenamefont {Muegge}\ \emph {et~al.}(1997)\citenamefont {Muegge},
  \citenamefont {Qi}, \citenamefont {Wand}, \citenamefont {Chu},\ and\
  \citenamefont {Warshel}}]{Muegge1997ET}%
  \BibitemOpen
  \bibfield  {author} {\bibinfo {author} {\bibfnamefont {I.}~\bibnamefont
  {Muegge}}, \bibinfo {author} {\bibfnamefont {P.~X.}\ \bibnamefont {Qi}},
  \bibinfo {author} {\bibfnamefont {A.~J.}\ \bibnamefont {Wand}}, \bibinfo
  {author} {\bibfnamefont {Z.~T.}\ \bibnamefont {Chu}}, \ and\ \bibinfo
  {author} {\bibfnamefont {A.}~\bibnamefont {Warshel}},\ }\href {\doibase
  10.1021/jp962478o} {\bibfield  {journal} {\bibinfo  {journal} {J.~Phys.
  Chem.~B}\ }\textbf {\bibinfo {volume} {101}},\ \bibinfo {pages} {825}
  (\bibinfo {year} {1997})}\BibitemShut {NoStop}%
\bibitem [{\citenamefont
  {Wolynes}(1987{\natexlab{b}})}]{Wolynes1987dissipation}%
  \BibitemOpen
  \bibfield  {author} {\bibinfo {author} {\bibfnamefont {P.~G.}\ \bibnamefont
  {Wolynes}},\ }\href {\doibase 10.1063/1.452146} {\bibfield  {journal}
  {\bibinfo  {journal} {J.~Chem. Phys.}\ }\textbf {\bibinfo {volume} {86}},\
  \bibinfo {pages} {1957} (\bibinfo {year} {1987}{\natexlab{b}})}\BibitemShut
  {NoStop}%
\bibitem [{\citenamefont {Kim}\ and\ \citenamefont
  {Kapral}(2006)}]{Kim2006bath}%
  \BibitemOpen
  \bibfield  {author} {\bibinfo {author} {\bibfnamefont {H.}~\bibnamefont
  {Kim}}\ and\ \bibinfo {author} {\bibfnamefont {R.}~\bibnamefont {Kapral}},\
  }\href {\doibase 10.1016/j.cplett.2006.03.025} {\bibfield  {journal}
  {\bibinfo  {journal} {Chem. Phys. Lett.}\ }\textbf {\bibinfo {volume}
  {423}},\ \bibinfo {pages} {76} (\bibinfo {year} {2006})}\BibitemShut
  {NoStop}%
\bibitem [{\citenamefont {Richardson}(2017)}]{formic}%
  \BibitemOpen
  \bibfield  {author} {\bibinfo {author} {\bibfnamefont {J.~O.}\ \bibnamefont
  {Richardson}},\ }\href {\doibase 10.1039/C6CP07808G} {\bibfield  {journal}
  {\bibinfo  {journal} {Phys. Chem. Chem. Phys.}\ }\textbf {\bibinfo {volume}
  {19}},\ \bibinfo {pages} {966} (\bibinfo {year} {2017})},\ \Eprint
  {http://arxiv.org/abs/1611.04816} {arXiv:1611.04816 [physics.chem-ph]}
  \BibitemShut {NoStop}%
\bibitem [{\citenamefont {Tiwari}\ and\ \citenamefont
  {Ensing}(2016)}]{Tiwari2016Faraday}%
  \BibitemOpen
  \bibfield  {author} {\bibinfo {author} {\bibfnamefont {A.}~\bibnamefont
  {Tiwari}}\ and\ \bibinfo {author} {\bibfnamefont {B.}~\bibnamefont
  {Ensing}},\ }\href {\doibase 10.1039/c6fd00132g} {\bibfield  {journal}
  {\bibinfo  {journal} {Faraday Discuss.}\ }\textbf {\bibinfo {volume} {195}},\
  \bibinfo {pages} {291} (\bibinfo {year} {2016})}\BibitemShut {NoStop}%
\bibitem [{\citenamefont {Hammes-Schiffer}(2015)}]{HammesSchiffer2015PCET}%
  \BibitemOpen
  \bibfield  {author} {\bibinfo {author} {\bibfnamefont {S.}~\bibnamefont
  {Hammes-Schiffer}},\ }\href {\doibase 10.1021/jacs.5b04087} {\bibfield
  {journal} {\bibinfo  {journal} {J. Am. Chem. Soc.}\ }\textbf {\bibinfo
  {volume} {137}},\ \bibinfo {pages} {8860} (\bibinfo {year}
  {2015})}\BibitemShut {NoStop}%
\bibitem [{\citenamefont {Hele}\ and\ \citenamefont
  {Althorpe}(2013)}]{Hele2013QTST}%
  \BibitemOpen
  \bibfield  {author} {\bibinfo {author} {\bibfnamefont {T.~J.~H.}\
  \bibnamefont {Hele}}\ and\ \bibinfo {author} {\bibfnamefont {S.~C.}\
  \bibnamefont {Althorpe}},\ }\href {\doibase 10.1063/1.4792697} {\bibfield
  {journal} {\bibinfo  {journal} {J.~Chem. Phys.}\ }\textbf {\bibinfo {volume}
  {138}},\ \bibinfo {pages} {084108} (\bibinfo {year} {2013})}\BibitemShut
  {NoStop}%
\bibitem [{\citenamefont {Shushkov}(2013)}]{Shushkov2013instanton}%
  \BibitemOpen
  \bibfield  {author} {\bibinfo {author} {\bibfnamefont {P.}~\bibnamefont
  {Shushkov}},\ }\href {\doibase 10.1063/1.4807706} {\bibfield  {journal}
  {\bibinfo  {journal} {J.~Chem. Phys.}\ }\textbf {\bibinfo {volume} {138}},\
  \bibinfo {pages} {224102} (\bibinfo {year} {2013})}\BibitemShut {NoStop}%
\bibitem [{\citenamefont {Habershon}\ \emph {et~al.}(2013)\citenamefont
  {Habershon}, \citenamefont {Manolopoulos}, \citenamefont {Markland},\ and\
  \citenamefont {Miller~III}}]{Habershon2013RPMDreview}%
  \BibitemOpen
  \bibfield  {author} {\bibinfo {author} {\bibfnamefont {S.}~\bibnamefont
  {Habershon}}, \bibinfo {author} {\bibfnamefont {D.~E.}\ \bibnamefont
  {Manolopoulos}}, \bibinfo {author} {\bibfnamefont {T.~E.}\ \bibnamefont
  {Markland}}, \ and\ \bibinfo {author} {\bibfnamefont {T.~F.}\ \bibnamefont
  {Miller~III}},\ }\href {\doibase 10.1146/annurev-physchem-040412-110122}
  {\bibfield  {journal} {\bibinfo  {journal} {Annu. Rev. Phys. Chem.}\ }\textbf
  {\bibinfo {volume} {64}},\ \bibinfo {pages} {387} (\bibinfo {year}
  {2013})}\BibitemShut {NoStop}%
\bibitem [{\citenamefont {Shushkov}, \citenamefont {Li},\ and\ \citenamefont
  {Tully}(2012)}]{Shushkov2012RPSH}%
  \BibitemOpen
  \bibfield  {author} {\bibinfo {author} {\bibfnamefont {P.}~\bibnamefont
  {Shushkov}}, \bibinfo {author} {\bibfnamefont {R.}~\bibnamefont {Li}}, \ and\
  \bibinfo {author} {\bibfnamefont {J.~C.}\ \bibnamefont {Tully}},\ }\href
  {\doibase 10.1063/1.4766449} {\bibfield  {journal} {\bibinfo  {journal}
  {J.~Chem. Phys.}\ }\textbf {\bibinfo {volume} {137}},\ \bibinfo {pages}
  {22A549} (\bibinfo {year} {2012})}\BibitemShut {NoStop}%
\bibitem [{\citenamefont {Richardson}\ and\ \citenamefont
  {Thoss}(2013)}]{mapping}%
  \BibitemOpen
  \bibfield  {author} {\bibinfo {author} {\bibfnamefont {J.~O.}\ \bibnamefont
  {Richardson}}\ and\ \bibinfo {author} {\bibfnamefont {M.}~\bibnamefont
  {Thoss}},\ }\href {\doibase 10.1063/1.4816124} {\bibfield  {journal}
  {\bibinfo  {journal} {J.~Chem. Phys.}\ }\textbf {\bibinfo {volume} {139}},\
  \bibinfo {pages} {031102} (\bibinfo {year} {2013})}\BibitemShut {NoStop}%
\bibitem [{\citenamefont {Richardson}\ \emph {et~al.}(2017)\citenamefont
  {Richardson}, \citenamefont {Meyer}, \citenamefont {Pleinert},\ and\
  \citenamefont {Thoss}}]{vibronic}%
  \BibitemOpen
  \bibfield  {author} {\bibinfo {author} {\bibfnamefont {J.~O.}\ \bibnamefont
  {Richardson}}, \bibinfo {author} {\bibfnamefont {P.}~\bibnamefont {Meyer}},
  \bibinfo {author} {\bibfnamefont {M.-O.}\ \bibnamefont {Pleinert}}, \ and\
  \bibinfo {author} {\bibfnamefont {M.}~\bibnamefont {Thoss}},\ }\href
  {\doibase 10.1016/j.chemphys.2016.09.036} {\bibfield  {journal} {\bibinfo
  {journal} {Chem. Phys.}\ }\textbf {\bibinfo {volume} {482}},\ \bibinfo
  {pages} {124} (\bibinfo {year} {2017})},\ \Eprint
  {http://arxiv.org/abs/1609.00644} {arXiv:1609.00644 [physics.chem-ph]}
  \BibitemShut {NoStop}%
\bibitem [{\citenamefont {Ananth}(2013)}]{Ananth2013MVRPMD}%
  \BibitemOpen
  \bibfield  {author} {\bibinfo {author} {\bibfnamefont {N.}~\bibnamefont
  {Ananth}},\ }\href {\doibase 10.1063/1.4821590} {\bibfield  {journal}
  {\bibinfo  {journal} {J.~Chem. Phys.}\ }\textbf {\bibinfo {volume} {139}},\
  \bibinfo {pages} {124102} (\bibinfo {year} {2013})}\BibitemShut {NoStop}%
\bibitem [{\citenamefont {Hele}\ and\ \citenamefont
  {Ananth}(2016)}]{Hele2016Faraday}%
  \BibitemOpen
  \bibfield  {author} {\bibinfo {author} {\bibfnamefont {T.~J.~H.}\
  \bibnamefont {Hele}}\ and\ \bibinfo {author} {\bibfnamefont {N.}~\bibnamefont
  {Ananth}},\ }\href {\doibase 10.1039/C6FD00106H} {\bibfield  {journal}
  {\bibinfo  {journal} {Faraday Discuss.}\ }\textbf {\bibinfo {volume} {195}},\
  \bibinfo {pages} {269} (\bibinfo {year} {2016})}\BibitemShut {NoStop}%
\bibitem [{\citenamefont {Duke}\ and\ \citenamefont
  {Ananth}(2016)}]{Duke2016Faraday}%
  \BibitemOpen
  \bibfield  {author} {\bibinfo {author} {\bibfnamefont {J.~R.}\ \bibnamefont
  {Duke}}\ and\ \bibinfo {author} {\bibfnamefont {N.}~\bibnamefont {Ananth}},\
  }\href {\doibase 10.1039/C6FD00123H} {\bibfield  {journal} {\bibinfo
  {journal} {Faraday Discuss.}\ }\textbf {\bibinfo {volume} {195}},\ \bibinfo
  {pages} {253} (\bibinfo {year} {2016})}\BibitemShut {NoStop}%
\bibitem [{\citenamefont {Chowdhury}\ and\ \citenamefont
  {Huo}()}]{Chowdhury2017CSRPMD}%
  \BibitemOpen
  \bibfield  {author} {\bibinfo {author} {\bibfnamefont {S.}~\bibnamefont
  {Chowdhury}}\ and\ \bibinfo {author} {\bibfnamefont {P.}~\bibnamefont
  {Huo}},\ }\href@noop {} {\ }\Eprint {http://arxiv.org/abs/1706.08403}
  {arXiv:1706.08403 [physics.chem-ph]} \BibitemShut {NoStop}%
\bibitem [{\citenamefont {Menzeleev}, \citenamefont {Bell},\ and\ \citenamefont
  {Miller~III}(2014)}]{Menzeleev2014kinetic}%
  \BibitemOpen
  \bibfield  {author} {\bibinfo {author} {\bibfnamefont {A.~R.}\ \bibnamefont
  {Menzeleev}}, \bibinfo {author} {\bibfnamefont {F.}~\bibnamefont {Bell}}, \
  and\ \bibinfo {author} {\bibfnamefont {T.~F.}\ \bibnamefont {Miller~III}},\
  }\href {\doibase 10.1063/1.4863919} {\bibfield  {journal} {\bibinfo
  {journal} {J.~Chem. Phys.}\ }\textbf {\bibinfo {volume} {140}},\ \bibinfo
  {pages} {064103} (\bibinfo {year} {2014})}\BibitemShut {NoStop}%
\bibitem [{\citenamefont {Friesner}, \citenamefont {Pettitt},\ and\
  \citenamefont {Jean}(1985)}]{Friesner1985Raman}%
  \BibitemOpen
  \bibfield  {author} {\bibinfo {author} {\bibfnamefont {R.}~\bibnamefont
  {Friesner}}, \bibinfo {author} {\bibfnamefont {M.}~\bibnamefont {Pettitt}}, \
  and\ \bibinfo {author} {\bibfnamefont {J.~M.}\ \bibnamefont {Jean}},\ }\href
  {\doibase 10.1063/1.448239} {\bibfield  {journal} {\bibinfo  {journal}
  {J.~Chem. Phys.}\ }\textbf {\bibinfo {volume} {82}},\ \bibinfo {pages} {2918}
  (\bibinfo {year} {1985})}\BibitemShut {NoStop}%
\bibitem [{\citenamefont {Balian}\ and\ \citenamefont
  {Brezin}(1969)}]{Balian1969Bogoliubov}%
  \BibitemOpen
  \bibfield  {author} {\bibinfo {author} {\bibfnamefont {R.}~\bibnamefont
  {Balian}}\ and\ \bibinfo {author} {\bibfnamefont {E.}~\bibnamefont
  {Brezin}},\ }\href {\doibase 10.1007/BF02710281} {\bibfield  {journal}
  {\bibinfo  {journal} {Nuovo Cimento B (1965--1970)}\ }\textbf {\bibinfo
  {volume} {64}},\ \bibinfo {pages} {37} (\bibinfo {year} {1969})}\BibitemShut
  {NoStop}%
\end{thebibliography}
\end{document}